\documentclass[twocolumn,times]{aastex62}

\received{2023}
\submitjournal{ApJ}

\shortauthors{Allen et al.}

\begin{document}

\title{High-resolution study of outflow activity and chemical environment of Chamaeleon-MMS1}

\correspondingauthor{Veronica Allen}
\email{allen@astro.rug.nl}

\author{Veronica Allen}

\affiliation{Kapteyn Astronomical Institute, University of Groningen\\
        9747 AD Groningen, the Netherlands}
\affiliation{NWO Veni Fellow}
\affiliation{NASA Postdoctoral Program fellow}

\author{Martin A. Cordiner}
\affiliation{NASA Goddard Space Flight Center \\
	8800 Greenbelt Road  \\
	Greenbelt, MD 20771, USA}
\affiliation{Institute for Astrophysics and Computational Sciences, The Catholic University of America \\
	Washington, DC 20064, USA}

\author{Gilles Adande}
\affiliation{NASA Postdoctoral Program fellow}
\affiliation{NASA Goddard Space Flight Center \\
	8800 Greenbelt Road  \\
	Greenbelt, MD 20771, USA}

\author{Steven B. Charnley}
\affiliation{NASA Goddard Space Flight Center \\
	8800 Greenbelt Road  \\
	Greenbelt, MD 20771, USA}

\author{Yi-Jehng Kuan}
\affiliation{Department of Earth Sciences, National Taiwan Normal University\\
	Taipei, 11677, Taiwan, ROC}
\affiliation{Institute of Astronomy and Astrophysics, Academia Sinica \\
	Taipei, 10617, Taiwan, ROC}

\author{Eva Wirstr\"{o}m}
\affiliation{Department of Space, Earth, and Environment, Chalmers University of Technology \\
	Onsala Space Observatory \\
	439 92 Onsala, Sweden}






\begin{abstract}
The earliest stages of low-mass star-formation are unclear, but it has been proposed that Chamaeleon-MMS1 is a candidate First Hydrostatic Core (FHSC), the transition stage between a prestellar and protostellar core. This work describes the local ($\sim$4000~AU) outflow activity associated with Chamaeleon-MMS1 and its effect on the surrounding material in order to determine the evolutionary state of this young low-mass source. We used the Atacama Large Millimeter/sub-millimeter Array (ALMA) to observe Chamaeleon-MMS1 at 220~GHz at high spatial ($\sim75$~AU) and spectral resolutions (0.1-0.3~km~s$^{-1}$). A low energy outflow is detected through its interaction with the surrounding cloud. The outflow consists of two components, a broad spectral feature ($\Delta$v$\sim$8~km~s$^{-1}$) to the northeast and narrow spectral features ($\Delta$v$\sim$1~km~s$^{-1}$) to both the northeast and southwest. The molecular tracers CS, formaldehyde (H$_2$CO), and methanol (CH$_3$OH) were used to analyze the effect of the outflow on the surrounding gas as well as determine the rotational temperature of this gas. The rotational temperature of H$_2$CO is calculated to be 40~K toward the continuum source with similarly low temperatures (10-75~K) toward clumps affected by the outflow. CH$_3$OH is only detected toward gas clumps located away from the continuum source, where the methanol is expected to have been released by the energy of the outflow through ice sputtering in shock waves. Chamaeleon-MMS1 shows outflow activity with the power source likely being a single precessing outflow. While molecular emission and high outflow speeds rule it out as a FHSC, the outflow is less energetic than outflows detected from other Class 0 objects. The inferred gas temperatures toward the continuum source are also relatively low, indicating that Cha-MMS1 is one of the youngest known sources.
\end{abstract}

\keywords{Interstellar Medium: Young Stellar Objects --- Star Formation --- individual objects: Cha-MMS1 --- stellar winds}


\section{Introduction} \label{sec:intro}

Understanding star formation is a central problem in astrophysics, with direct implications for our understanding of how galaxies evolve and how planets and life form. Stars form in collapsing cores, corresponding to the densest regions within molecular clouds, but the precise details of this process are still far from being completely understood, particularly during the initial collapse period. The numerical hydrodynamic simulations of \citet{Larson1969}, more recently by \citet{Masunaga2000}, \citet{Machida2008}, and \citet{Tomida2010}, show that the collapse of a molecular cloud is arrested by thermal pressure once the density (\textit{n}) becomes sufficiently great that cooling by dust radiation is inhibited (at \textit{n} $\gtrsim$ 10$^{11}$ cm$^{-3}$). At this point, a quasi-hydrostatic object is formed, known as the First Hydrostatic Core (FHSC). The FHSC is calculated to persist in an approximately adiabatic state for $\sim$10$^3$ yr, which is a sufficiently short time to make it a rare object in nearby star-forming regions. Observational signatures of the FHSC include a low luminosity ($\sim$0.01-0.1L$_{\odot}$), and an approximately thermal spectral energy distribution that peaks in the far-infrared/sub-mm range \citep{Saigo2011}. A low luminosity source is generally considered to be a FHSC candidate if it shows little to no emission at wavelengths $< 70\mu$m. Calculations show that the FHSC should also produce a slow molecular outflow, which extends up to a distance of a few hundred AU from the core \citep{Machida2008,Tomida2010}. Typically, slow uncollimated outflows are expected, but \citet{Price2012} demonstrated that collimated jets with velocities up to 7~km~s$^{-1}$ are possible during the FHSC phase. 

\par Several FHSC candidates are currently identified. These include L1451-mm \citep{Pineda2011}; NGC 1333 - IRAS 4C \citep{Evgenia2016}; L1535-NE/MC35-mm \citep{Fujishiro2020}; Oph A N6 and Oph A SM1 \citep{Friesen2018}; and G208.89-20. \citep{Dutta2022}; and perhaps Per-Bolo 45 \citet{MariaJose2020}. Distinguishing between an FHSC and a very low luminosity Class 0 protostar (known as a VeLLO) has proven difficult, in part due to a lack of detailed spectro-spatial information on outflow properties. VeLLOs appear to be young, low-mass protostars undergoing episodic accretion, and a number of these were identified by the c2d Spitzer survey of \citet{Dunham2008}. The list of very low luminosity protostars includes IRAS 15398-3359 \citep{Okoda2020}; L673-7 IRS \citep{Dunham2011}; L1014 IRS \citep{Dunham2011}; L1148 IRS \citep{Dunham2011}; GSS30, SM1N, B2-A7 and IRAS 16267-2417 in Oph A \citep{Friesen2018}, as well as sources formerly identified as FHSC candidates: B1b-N \citep{Marcelino2018} and B1b-S \citep{Hirano2019}; Per-Bolo 58, L1448 IRS 2E and L1448 IRS 2E \citep{MariaJose2020}; CB17-MMS \citep{Spear2021}. The physics, chemistry, and observational signatures of FHSCs and VeLLOs have been discussed by \citet{Omukai2007}, \citet{Lee2007}, and \citet{Young2019}. Specifically, CS emission was modeled to exist only in the central few AU around a late-stage FHSC.

\par Chamaeleon-MMS1 (Cha-MMS1) is a young protostar located at a distance of 192~pc \citep{Dzib2018} with a systemic velocity of 4.5~km~s$^{-1}$ \citep{Belloche2006} and was first detected in 1.3~mm emission by \citet{Reipurth1996}. The central object is deeply embedded inside a dense molecular cloud \citep{Cordiner2012} and has been classified by \citet{Belloche2006,Belloche2011} as either an extremely young VeLLO or an FHSC with an internal luminosity between 0.08 and 0.18 L$_{\sun}$ \citep{Tsitali}. \citet{Belloche2006} did not find any evidence of CO outflow activity from Cha-MMS1 using single-dish (APEX) observations, and the studies by \citet{Hiramatsu2007,Ladd2011} determined that the nearby HH 49/50 outflow is driven by the more evolved Class I protostar Ced 110 IRS 4. \citet{Ladd2011} determined from single-dish CO observations that the IRS 4 outflow is deflected around (or grazes the edge of) the 100$''$-diameter dense clump that contains the collapsing protostar Cha-MMS1. \citet{Vaisala2014} analyzed NH$_3$ and continuum emission toward Cha-MMS1 using the Australia Telescope Compact Array (ATCA) and concluded that a warm central source was within, but it was at the Class 0 stage at the latest. Recently, \citet{Busch2020} published an analysis of the CO(3-2) emission tracing outflow activity associated with Cha-MMS1. The outflow in that article is qualitatively similar to the CO(2-1) emission in this work, but we also present the outflow activity as traced by CS, H$_{2}$CO, and CH$_{3}$OH, showing different structures than those traced by the CO observation. Additionally, analysis of multiple FHSC candidates (including Cha-MMS1) using dense envelope tracers was published by \citet{MariaJose2020}. Both studies conclude that Cha-MMS1 is at a very young evolutionary state. 

\par In this paper, we present the high-angular resolution analysis of the outflow activity associated with Cha-MMS1 and their effect on the surrounding material. In $\S$\ref{sec:obs}, we give details of the ALMA observations and data reduction. $\S$\ref{sec:results} contains the details of five detected species including isotopologs (CO, $^{13}$CO, CS, H$_{2}$CO and CH$_{3}$OH), analysis of the outflow mechanical properties, and determination of the rotational temperature of various sub-regions in this source. $\S$\ref{sec:discuss}, we discuss the implications of our results, and in $\S$\ref{sec:conc} we present our conclusions regarding the nature of this extremely young protostellar object.

\begin{figure}[!h]
	\centering
	\includegraphics[width=\hsize]{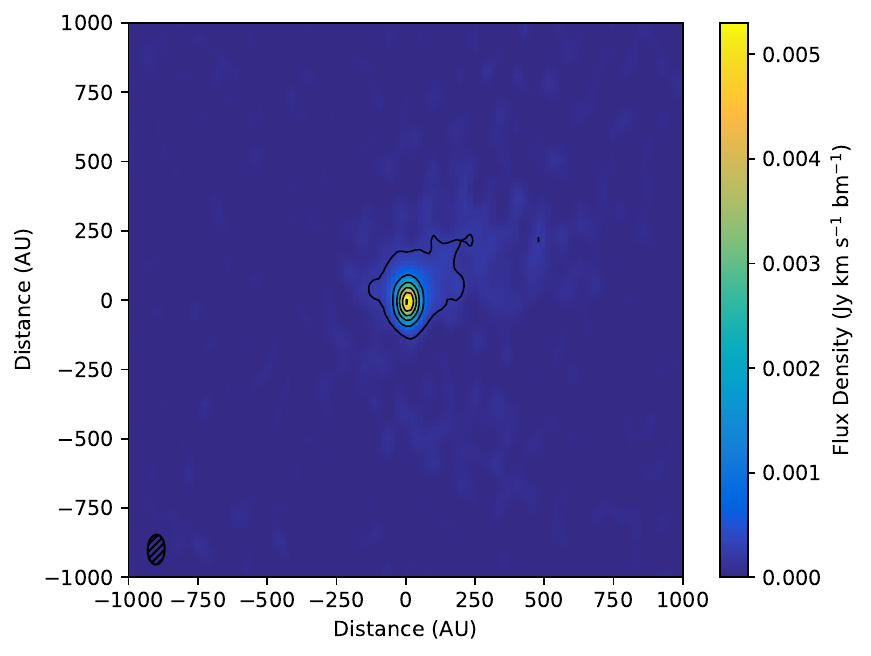}
	\caption{220~GHz continuum map of Cha-MMS1. Contours start at 3$\sigma$ (with an rms of 0.21 mJy/beam) continue at 1.05, 2.10, 3.15, 4.20, and 5.25 mJy/beam with a peak of 6.48 mJy/beam. The beam size is 0.72$''$x0.41$''$.}
	\label{Contmap}
\end{figure}

\begin{table}[h!]
	\centering
	\caption{Targeted spectral lines (E$_{\mathrm{up}}$ and A$_{\mathrm{ij}}$ from CDMS)}
	\label{linesTable}
	\begin{tabular}{ccccc}
		
		\hline\hline
		Species & Transition & Frequency & E$_{\mathrm{up}}$ & log (A$_{\mathrm{ij}}$) \\
         & & (GHz) & (K) & \\
		\hline
		CO & 2-1 & 230.5380 & 16.6 & -6.16 \\
		$^{13}$CO & 2-1 & 220.3986 & 15.9 & -6.22 \\
		CS & 5-4 & 244.9355 & 35.3 & -3.53 \\
		H$_2$CO & 3$_{0,3}$-2$_{2,1}$ & 218.2221 & 20.9 & -3.55 \\
		H$_2$CO & 3$_{2,2}$-2$_{2,1}$ & 218.4756 & 68.1 & -3.80 \\
		H$_2$CO & 3$_{1,2}$-2$_{1,1}$ & 225.6977 & 33.4 & -3.56 \\
		CH$_3$OH & 5$_{3,2}$-4$_{3,1}$ & 218.4400 & 45.5 & -4.33 \\
		CH$_3$OH & 5$_{-1,5}$-4$_{-1,4}$ & 241.7672 & 40.4 & -4.24 \\
		CH$_3$OH & 5$_{0,5}$-4$_{0,4}$++ & 241.7914 & 34.8 & -4.22 \\
		HC$_3$N & 24-23 & 218.3247 & 131.0 & -3.08 \\
		\hline
	\end{tabular}
\end{table}

\begin{figure}[!h]
	\centering
	\includegraphics[width=\hsize]{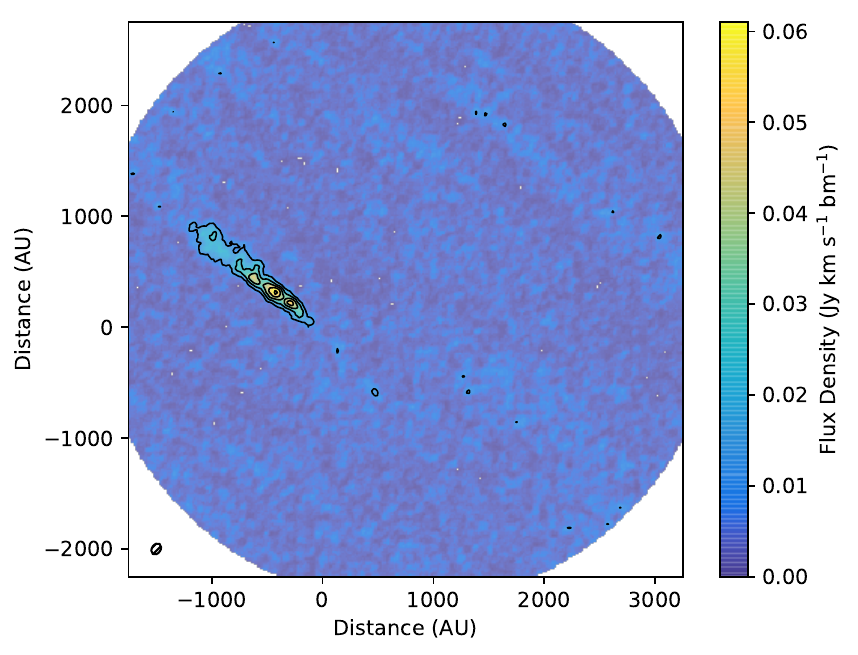}
	\caption{Primary beam corrected integrated intensity map (between 8.7 and 17.1~km~s$^{-1}$) of broad component of CO outflow. Contours start at 3$\sigma$ (rms of 12 mJy/beam~km~s$^{-1}$) continue in intervals of 3$\sigma$ to a peak of 61 mJy/beam~km~s$^{-1}$. The beam size is 0.72$''$x0.52$''$.}
	\label{CObroadred}
\end{figure}

\begin{figure*}[!ht]
	\begin{minipage}{0.5\textwidth}
		\centering
		\includegraphics[width=\textwidth]{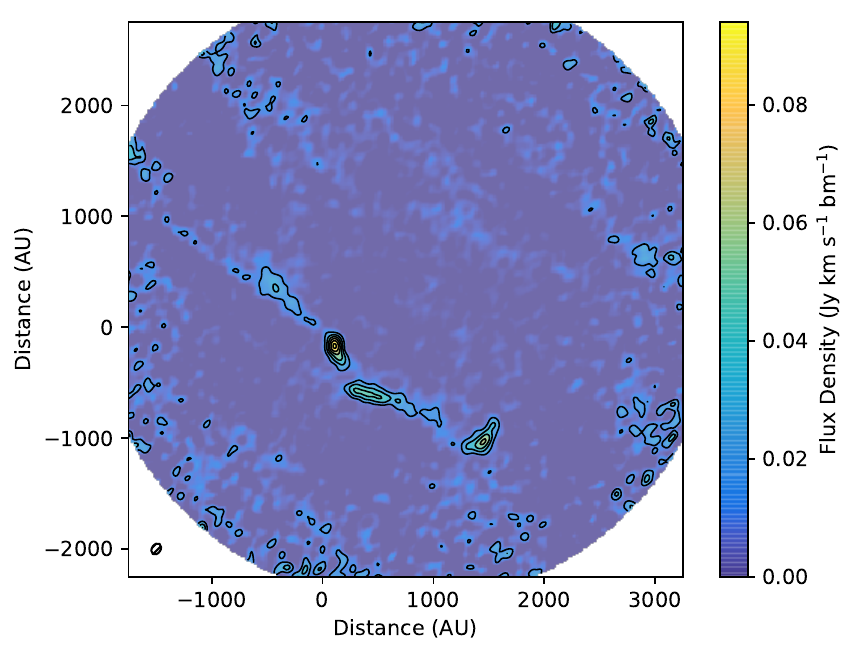}
	\end{minipage}
	\begin{minipage}{0.5\textwidth}
		\centering
		\includegraphics[width=\textwidth]{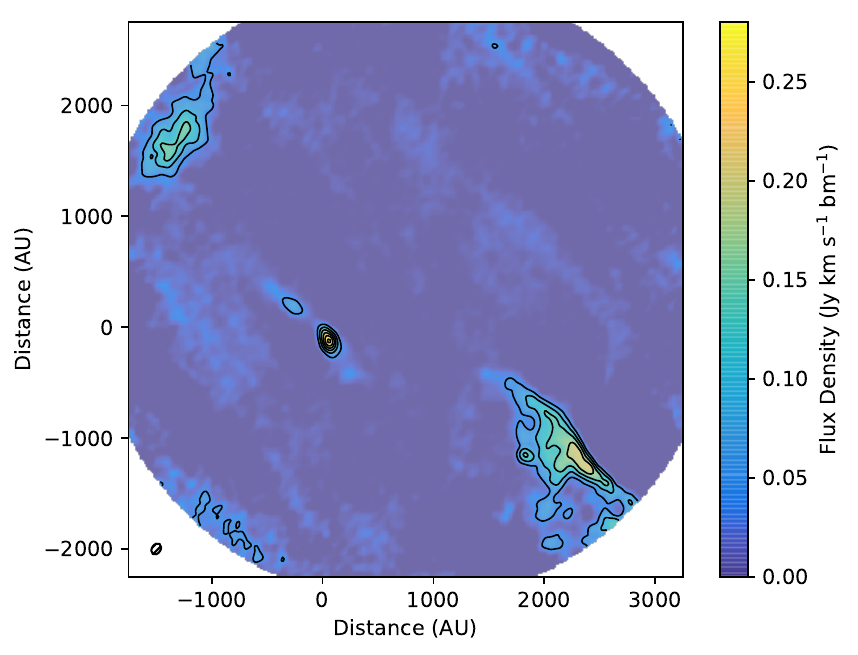}
	\end{minipage}
	\caption{Primary beam corrected integrated intensity maps for $^{12}$CO narrow emission features.  The continuum peak is located at 0,0 AU. The beam size is 0.72$''$x0.52$''$.
		\textit{(left)} Integrated intensity map (between 5.7 and 8.7~km~s$^{-1}$) of narrow red-shifted component of CO outflow. Contours start at 3$\sigma$ (rms of 15 mJy/beam~km~s$^{-1}$) continue in intervals of 3$\sigma$ to a peak of 94 mJy/beam~km~s$^{-1}$.
		\textit{(right)} Integrated intensity map (between 1.8 and 4.0~km~s$^{-1}$) of narrow blue-shifted lobes of CO outflow. Contours start at 3$\sigma$ (rms of 45 mJy/beam~km~s$^{-1}$) continue in intervals of 3$\sigma$ up to a peak of 277 mJy/beam~km~s$^{-1}$.}
	\label{COcontours}
\end{figure*}

\section{Observations and data reduction} \label{sec:obs}
Cha-MMS1 was observed using the Atacama Large Millimeter/sub-millimeter Array (ALMA) on 16 July 2014 and 6 and 7 June 2015 using 26 antennas on 16 July, and 37-39 antennas on 6 and 7 June (project code 2013.1.01113.S) with an angular resolution of $\sim0.7''$ and spectral resolutions of $\sim$0.08~km~s$^{-1}$ in three spectral windows (covering $^{12}$CO, $^{13}$CO and CS), 0.17~km~s$^{-1}$ for H$_2$CO (218222~MHz), and 0.3~km~s$^{-1}$ in the other four spectral windows. The calibrators used were Callisto, Ganymede, J0635-7516, J1058-8003, J1208-7809, J1107-4449, and J0538-4405. The pointing position of our ALMA observations was the millimeter position reported in \citet{Reipurth1996} rather than the Spitzer IR peak, therefore the source is consistently offset in our maps. Data reduction was performed with the CASA package (version 5.4.0). Corresponding spectral windows from different observation dates were concatenated and the continuum was subtracted. Imaging was done using the tclean process with a Hogbom deconvolver and Briggs weighting with a robust parameter of 0.5 and all images are primary beam corrected. The RMS noise for line-free channels was typically 2-4 mJy.


\begin{figure*}[!ht]
	\begin{minipage}{0.5\textwidth}
		\centering
		\includegraphics[width=\textwidth]{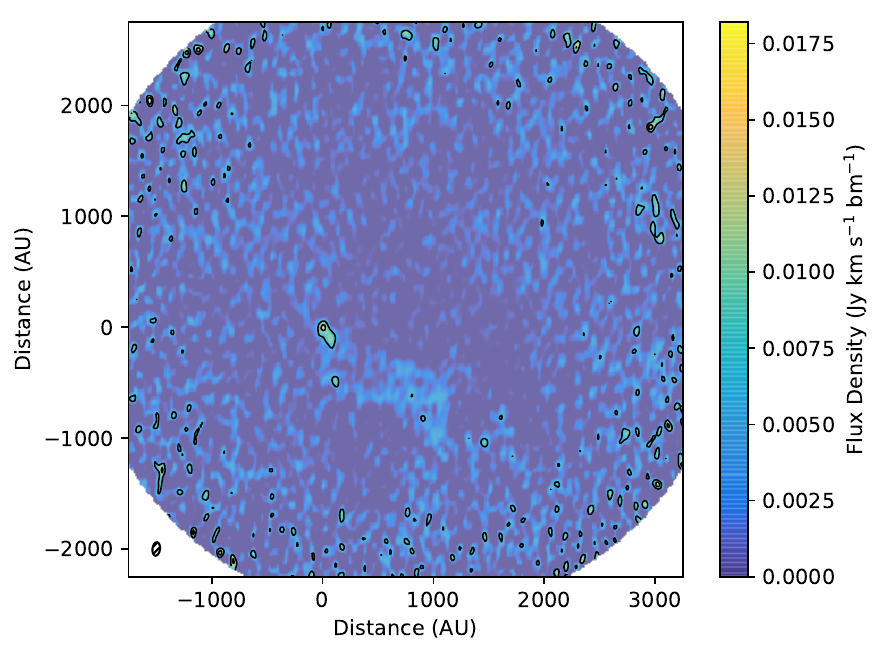}
	\end{minipage}
	\begin{minipage}{0.5\textwidth}
		\centering
		\includegraphics[width=\textwidth]{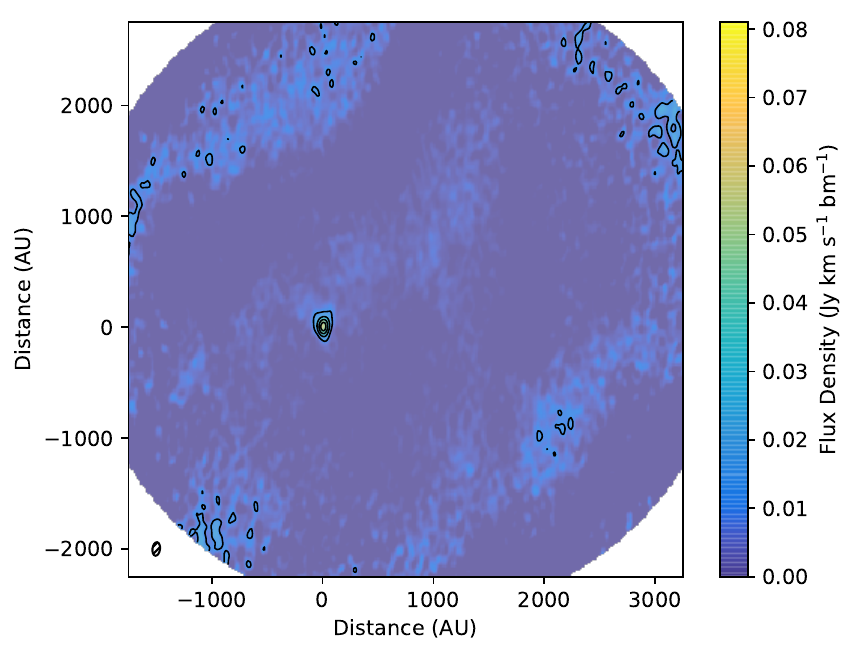}
	\end{minipage}
	\caption{Primary beam corrected integrated intensity maps for $^{13}$CO emission features.  The continuum peak is located at 0,0 AU. The beam size is 0.82$''$x0.45$''$
		\textit{(left)} Integrated intensity map (between 5.1 and 6.0~km~s$^{-1}$) of red component of $^{13}$CO outflow. Contours start at 3$\sigma$ (rms of 3.1 mJy/beam~km~s$^{-1}$) continue in intervals of 1$\sigma$ to a peak of 18.1 mJy/beam~km~s$^{-1}$.
		\textit{(right)} Integrated intensity map (between 3.5 and 4.4~km~s$^{-1}$) of blue component of $^{13}$CO outflow. Contours start at 3$\sigma$ (rms of 12 mJy/beam~km~s$^{-1}$) continue in intervals of 3$\sigma$ to a peak of 80.6 mJy/beam~km~s$^{-1}$.}
	\label{13COnarrowred}
\end{figure*}

\section{Results} \label{sec:results}
\subsection{Continuum and Line Detection}
Figure~\ref{Contmap} shows the compact, unresolved continuum emission at 220~GHz detected toward Cha-MMS 1. Our observations targeted several molecular species expected in very young low-mass star-forming regions summarized in Table~\ref{linesTable}. $^{12}$CO emission is resolved out on large scales, but a weak spectral feature covering the broad velocity range of 8.7-17.1~km~s$^{-1}$ is detected in $^{12}$CO to the northeast of the continuum source (\S~\ref{broadoutflow}, Figure~\ref{CObroadred}). Moment maps were also made of narrow spectral features in $^{12}$CO and $^{13}$CO (\S~\ref{narrowoutflow}, Figures~\ref{COcontours} and \ref{13COnarrowred}). There is self-absorption in both $^{12}$CO and $^{13}$CO with no signal for each in the ranges 4.0-5.7~km~s$^{-1}$ and 4.4-5.1~km~s$^{-1}$, respectively. $^{13}$CO shows strong emission towards the continuum source and weak emission (less than 3$\sigma$) cospatial with the narrow red $^{12}$CO emission southwest of the continuum source.
CS emission is clumpy and, on larger spatial scales, the CS clumps seem to trace out a cone (\S~\ref{CS}, Figure~\ref{CScontours}).  Within 1.5$''$ (288~AU) of the continuum source spectral emission of CS shows shows an hourglass-like morphology.
Three H$_2$CO lines were detected (\S~\ref{h2cosec}, Figure~\ref{H2COcontours}) showing emission in different regions, most of which correspond with the locations of CS emission.  
Three weak CH$_3$OH transitions were detected at 5- to 10$\sigma$ RMS levels (\S~\ref{ch3oh}) coinciding with the H$_2$CO emission that is not coincident with the Cha-MMS1 core.
The targeted HC$_3$N transition was not detected.

\begin{figure*}[!ht]
	\begin{minipage}{0.5\textwidth}
		\centering
		\includegraphics[width=\textwidth]{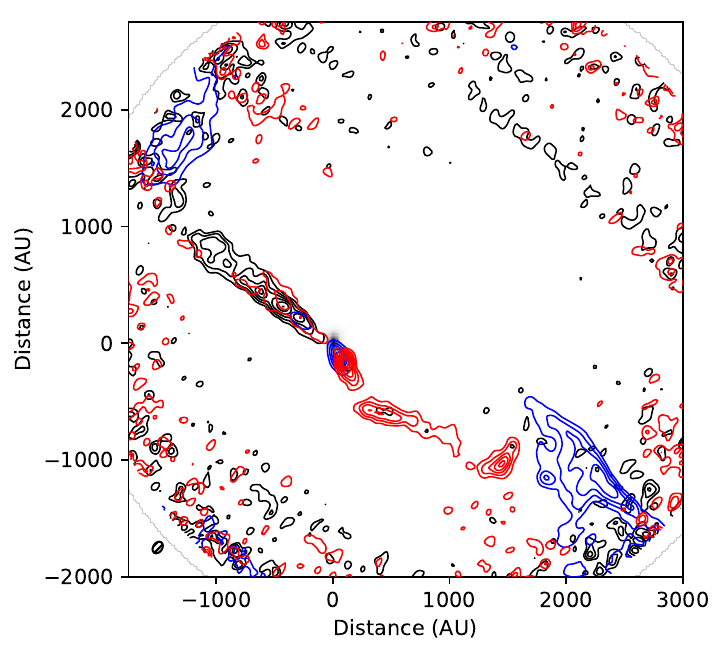}
	\end{minipage}
	\begin{minipage}{0.5\textwidth}
		\centering
		\includegraphics[width=\textwidth]{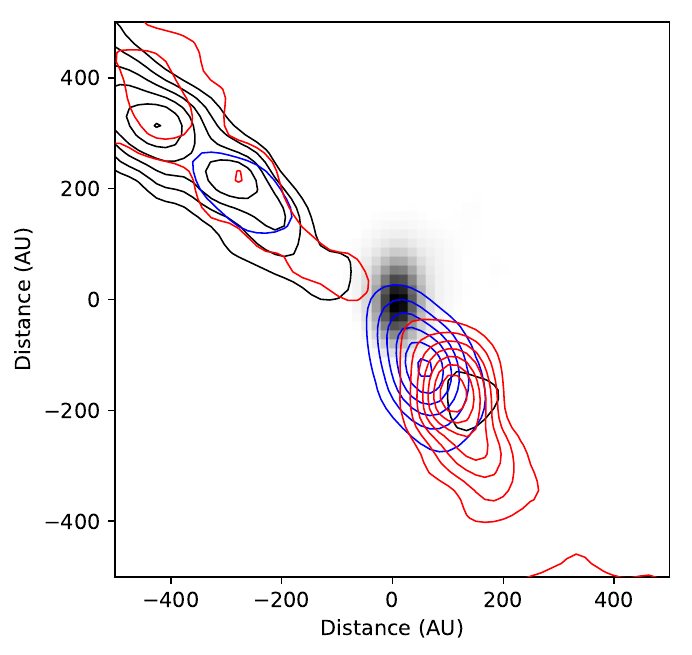}
	\end{minipage}
	\caption{Primary beam corrected integrated intensity image of the $^{12}$CO emission over the red-shifted, narrow velocity range from 5.7 to 8.7~km~s$^{-1}$ (red contours) and the blue-shifted, narrow velocity range from 3.5 to 4.4~km~s$^{-1}$ (blue contours). The $^{12}$CO emission of the red-shifted, broad velocity range from 8.7 to 17.1~km~s$^{-1}$ is shown in black contours and all are overlaid on the continuum emission of Cha-MMS1 (greyscale). Contours are as in Figures~\ref{CObroadred} and \ref{COcontours}. The \textbf{\textit{left}} figure shows the full extent of the emission and the \textbf{\textit{right}} figure shows the detail of the inner 1000 AU. The beam size is 0.72$''$x0.52$''$.}
	\label{COallcontours}
\end{figure*}

\subsection{Outflow activity}
The emission detected in $^{12}$CO was used to determine the properties of outflow activity associated with this source. Both the red and blue lobes of the narrow outflow component are located toward the northeast and southwest of the continuum source with an additional broad red component appearing to the northeast.

\subsubsection{Broad red outflow component}
\label{broadoutflow}
A weak, broad red-shifted outflow is detected to the northeast of the continuum source in $^{12}$CO (Figure~\ref{CObroadred}). It is quite collimated with a collimation factor of 5.5. A Gaussian fit made to this component of the  $^{12}$CO emission shows that the center is at 10.9$\pm$0.2~km~s$^{-1}$ and the full-width half-maximum (FWHM) is 6.7$\pm$0.4~km~s$^{-1}$.  The blue counterpart to the southwest is not nearly as broad or collimated (see Figure~\ref{COcontours} right). A Gaussian fit of the southern blue component has a central velocity of 3.40$\pm$0.01~km~s$^{-1}$ and a FWHM of 0.77$\pm$0.03~km~s$^{-1}$.  This much narrower spectral line may not be the opposite component of the broad red outflow, but no broader blue component was detected.  No broad component is detected in $^{13}$CO (Figure~\ref{13COnarrowred}).

\subsubsection{Narrow bipolar outflow}
\label{narrowoutflow}
The left frames of Figures~\ref{COcontours} and \ref{13COnarrowred} show $^{12}$CO and $^{13}$CO integrated intensity maps (between 5.7 and 8.7~km~s$^{-1}$ and 5.1 and 6.0~km~s$^{-1}$, respectively) of spectrally narrow red-shifted outflow activity to the south of the continuum source which coincide in an 'S'-like shape.  The blue components of $^{12}$CO (Figure~\ref{COcontours} right) are both to the northeast and southwest at the edges of the field-of-view. These blue components appear to be the ends of the outflow, but as we do not have a larger field-of-view, we cannot confirm. The line profiles of the $^{12}$CO components are narrow at 1.4$\pm$0.1~km~s$^{-1}$ (southern red) and 0.72$\pm$0.07~km~s$^{-1}$ (northern blue) with central peaks near v$_{LSR}$ of Cha-MMS1 at 6.94$\pm$0.04~km~s$^{-1}$ and 3.55$\pm$0.03~km~s$^{-1}$.  The spectral lines for $^{13}$CO are even narrower at 0.34$\pm$0.02 and 0.35$\pm$0.01 with central peaks at 5.47$\pm$0.01 and 3.565$\pm$0.006 for the red and blue lobes, respectively. All three components of the $^{12}$CO outflow -- broad red, narrow red, and narrow blue -- are shown in Figure~\ref{COallcontours}. Details of the spectra across the region can be found in Appendix A in Figures~\ref{cospec1} and \ref{cospec2}.

\begin{table*}[!ht]
	\centering
	\caption{Calculated mechanical outflow properties from CO emission (errors in parentheses) assuming $\tau$=5. V$_{\mathrm{max}}$ is the observed maximum velocity wrt V$_{\mathrm{LSR}}$. Velocity dependent properties are calculated assuming an inclination of 80$^\circ$. V$_{\mathrm{dep}}$ is the deprojected maximum velocity taking into account the outflow direction. Derived value ranged based on the trends in \citet{Wu2004} for a young star with a L$_{\mathrm{bol}}$ of 0.08-0.18 are started in the column "Calculated" with non-derivable velocity-based properties marked with $\dagger$.}
	\label{outflowTable}
	\begin{tabular}{cccccc}
		\hline\hline
		& Northern Red & Southern Red & Northern Blue & Southern Blue & Calculated \\
		\hline
		V$_{\mathrm{max}}$ (km s$^{-1}$)                & 12.4 (0.1)     & 4.0 (0.1)       & 3.2 (0.1) & 2.9 (0.1)  & $\dagger$    \\
		V$_{\mathrm{dep}}$ (km s$^{-1}$)            & 71.8 (0.6)   	& 23.4 (0.6)	& 18.6 (0.6) & 16.8 (0.6) & $\dagger$  \\
		t$_{\mathrm{kin}}$ (yr)                             & 130 (10)      & 490 (20)      &  720 (20)  & 800 (30)   & $\dagger$   \\
		Outflow mass (M$_{\sun}$)                        & 5.2 (1.0)$\times10^{-5}$  & 7.4 (1.7)$\times10^{-5}$   & 4.7 (0.6)$\times10^{-5}$ & 2.1 (0.1)$\times10^{-4}$  & 2.2-3.5$\times10^{-2}$ \\
		Mass-loss rate (M$_{\sun}$ yr$^{-1}$)         & 4.1 (1.2)$\times10^{-7}$  & 1.5 (0.4)$\times10^{-7}$  & 3.3 (0.7)$\times10^{-7}$ & 3.1 (0.3)$\times10^{-7}$  &  $\dagger$ \\
		Momentum (M$_{\sun}$~km~s$^{-1}$)            & 3.7 (0.8)$\times10^{-3}$  & 1.7 (0.4)$\times10^{-3}$   & 9.9 (1.2)$\times10^{-4}$ & 9.4 (0.7)$\times10^{-3}$  & $\dagger$ \\
		Kinetic Energy (erg)                                   & 2.6 (0.5)$\times10^{42}$  & 4.0 (0.9)$\times10^{41}$  & 1.6 (1.9)$\times10^{41}$ & 5.9 (0.4)$\times10^{41}$  & $\dagger$ \\
		Mechanical Luminosity (L$_{\sun}$)               & 1.1 (2.7)$\times10^{-1}$  & 5.7 (3.1)$\times10^{-3}$   & 2.3 (0.8)$\times10^{-3}$ & 2.4 (0.2)$\times10^{-2}$  & 2.2-3.6$\times10^{-3}$ \\
		Momentum rate (M$_{\sun}$~km~s$^{-1}$ yr$^{-1}$) & 2.9 (0.7)$\times10^{-5}$  & 3.6 (1.4)$\times10^{-6}$  & 1.4 (0.1)$\times10^{-6}$ & 1.1 (0.1)$\times10^{-5}$  & 2.2-4.0$\times10^{-6}$ \\
		\hline
	\end{tabular}
\end{table*}

\subsubsection{Mechanical outflow properties}
We used the approach described in \citet{AlvaroOutflows} and \citet{AnaOutflows} to calculate the mechanical outflow properties.  The measured properties for each outflow lobe are: the difference between the maximum velocity and systemic velocity (v$_{max}$), the length of the major axis (at the 5$\sigma$ contour), the area of the outflow (as an ellipse), and the integrated intensity of the emission between v$_{max}$ and 2$\sigma$ from the line peak ($\int$T$_{mb}$dv).  The CO column density was determined assuming optically thin emission as in \citet{Goldsmith1999} equations (10) and (19) assuming excitation temperatures of 10, 20, and 30~K. The calculated CO column densities varied by a factor of 3-4 between the different temperatures, so we used the average for calculating the hydrogen column density (N$_{H2}$) assuming a CO abundance of 10$^{-4}$, as is typical \citep{Bolatto2013}. Because the CO emission showed strong self-absorption at velocities near the velocity of the system (v$_{\mathrm{LSR}}$=4.9~km~s$^{-1}$), the CO intensities were corrected assuming an $\tau$ value of 5.
\par The other calculated properties are: outflow mass (calculated using the area of the outflow lobe and the mass of the hydrogen from the calculated H$_2$ column density), the dynamical or kinetic time -- t$_{kin}$ (the length of the outflow lobe divided by v$_{max}$), kinetic energy (outflow mass times 0.5 v$_{max}^2$), momentum (outflow mass times v$_{max}$), momentum rate -- which is also called the force (momentum/t$_{dyn}$), mass-loss rate (outflow mass/t$_{dyn}$), and mechanical luminosity (kinetic energy/t$_{dyn}$). As the direction of the outflow is near the plane of the sky, we applied the inclination correction factors from \citet{Ana2010} assuming an inclination of 80$^{\circ}$. We determined this inclination geometrically taking into account the opening angle of the northern red outflow and the presence of red and blue components in each direction.

\par The mechanical outflow properties are summarized in Table~\ref{outflowTable} for the detected outflow lobes. The fast broad outflow component (northern red) has a maximum relative velocity of $\sim$ 12~km s$^{-1}$ (deprojected to $\sim$ 70~km~s$^{-1}$). The narrow components have a maximum velocity of $\sim$4~km~s$^{-1}$ (deprojected to $\sim$20~km~s$^{-1}$). Using the trends outlined in \citet{Wu2004}, we calculated the momentum rate, outflow mass, and mechanical luminosity expected from a young star with a L$_{\mathrm{bol}}$ of 0.08-0.18. The momentum rate would be 2.2-4.0 $\times10^{-6}$ M$_{\sun}$~km~s$^{-1}$ yr$^{-1}$, the outflow mass 2.2-3.5 $\times10^{-2}$ M$_{\sun}$, and the mechanical luminosity 2.2-3.6 $\times10^{-3}$ L$_{\sun}$.
The outflows described in this paper have a mass 100-1000 times lower, and a momentum rate and mechanical luminosity comparable to the calculated value.

\subsection{CS}
\label{CS}
\begin{figure}[!bh]
	\centering
	\includegraphics[width=\hsize]{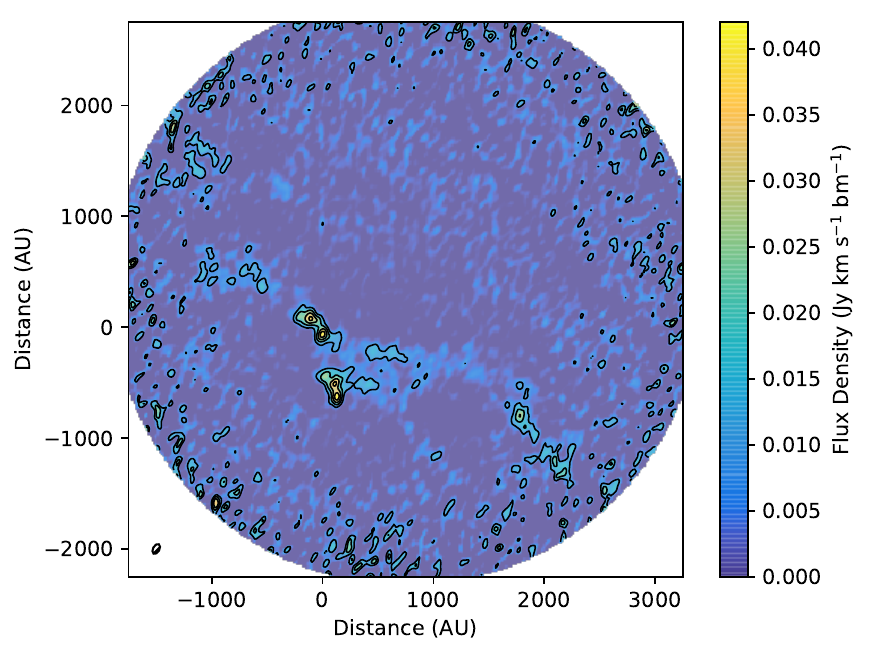}
	\caption{Primary beam corrected integrated intensity map (between 3.7 and 5.5~km~s$^{-1}$) of CS emission. Contours start at 3$\sigma$ (rms of 9 mJy/beam~km~s$^{-1}$) continue in intervals of 3$\sigma$ to a peak of 42.2 mJy/beam~km~s$^{-1}$. The beam size is 0.65$''$x0.39$''$.}
	\label{CScontours}
\end{figure}

\begin{figure*}[!hbt]
	\begin{minipage}{0.49\textwidth}
		\centering
		\includegraphics[width=\textwidth]{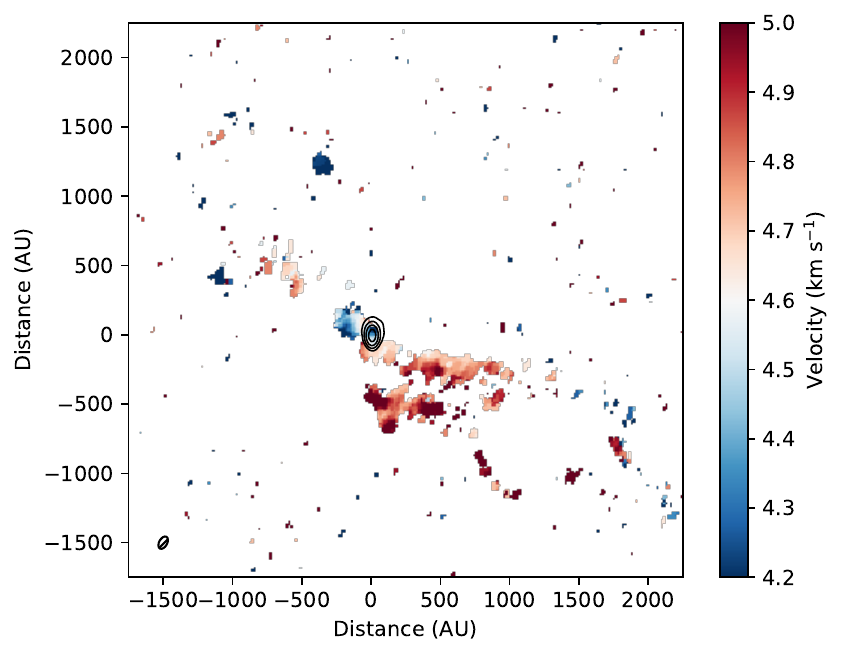}
	\end{minipage}
	\begin{minipage}{0.49\textwidth}
		\centering
		\includegraphics[width=\textwidth]{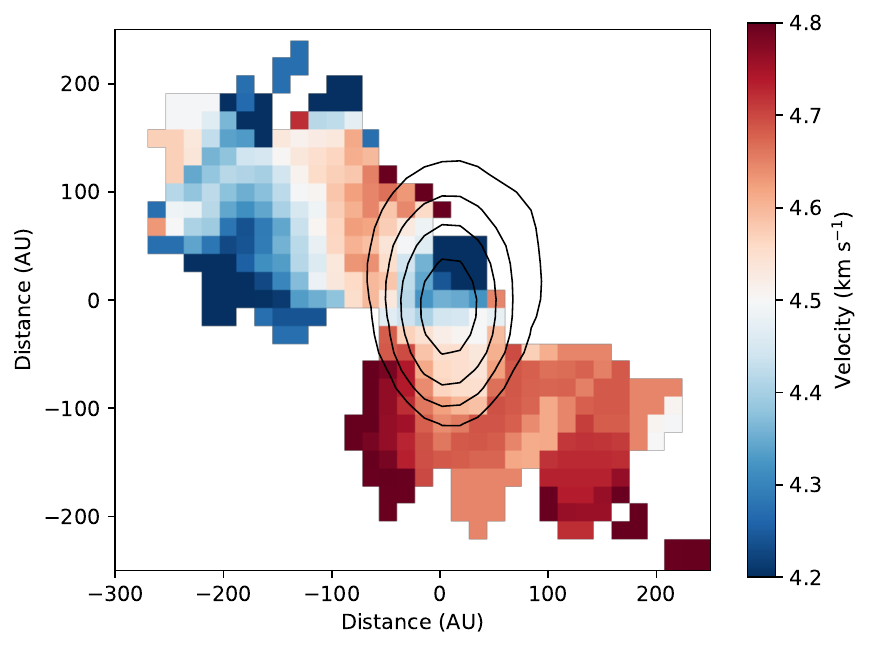}
	\end{minipage}
	\caption{Colored pixels correspond to the velocity at each point. The black contours show the continuum source. \textbf{\textit{Left:}} First moment map (velocity) of CS emission across the field-of-view. \textbf{\textit{Right:}} As in Left, but focused on emission near the continuum source.}
	\label{CSmom1}
\end{figure*}

Figure~\ref{CScontours} shows CS emission tracing a wide cone on larger scales, indicating the interaction between multiple outflows and the surroundings or a steep precessing angle of a single outflow. Alternatively, this morphology could be due to a spatially wider and low-velocity component tracing the cavity walls, as is commonly found towards other protostellar outflows. On smaller scales, the CS emission is hourglass-shaped and the first moment (velocity) maps (Figure~\ref{CSmom1}) show inverse velocity gradients immediately above and below the continuum source.

\subsection{H$_2$CO emission}
\label{h2cosec}
\begin{figure}[hb!]
	\centering
	\includegraphics[width=\hsize]{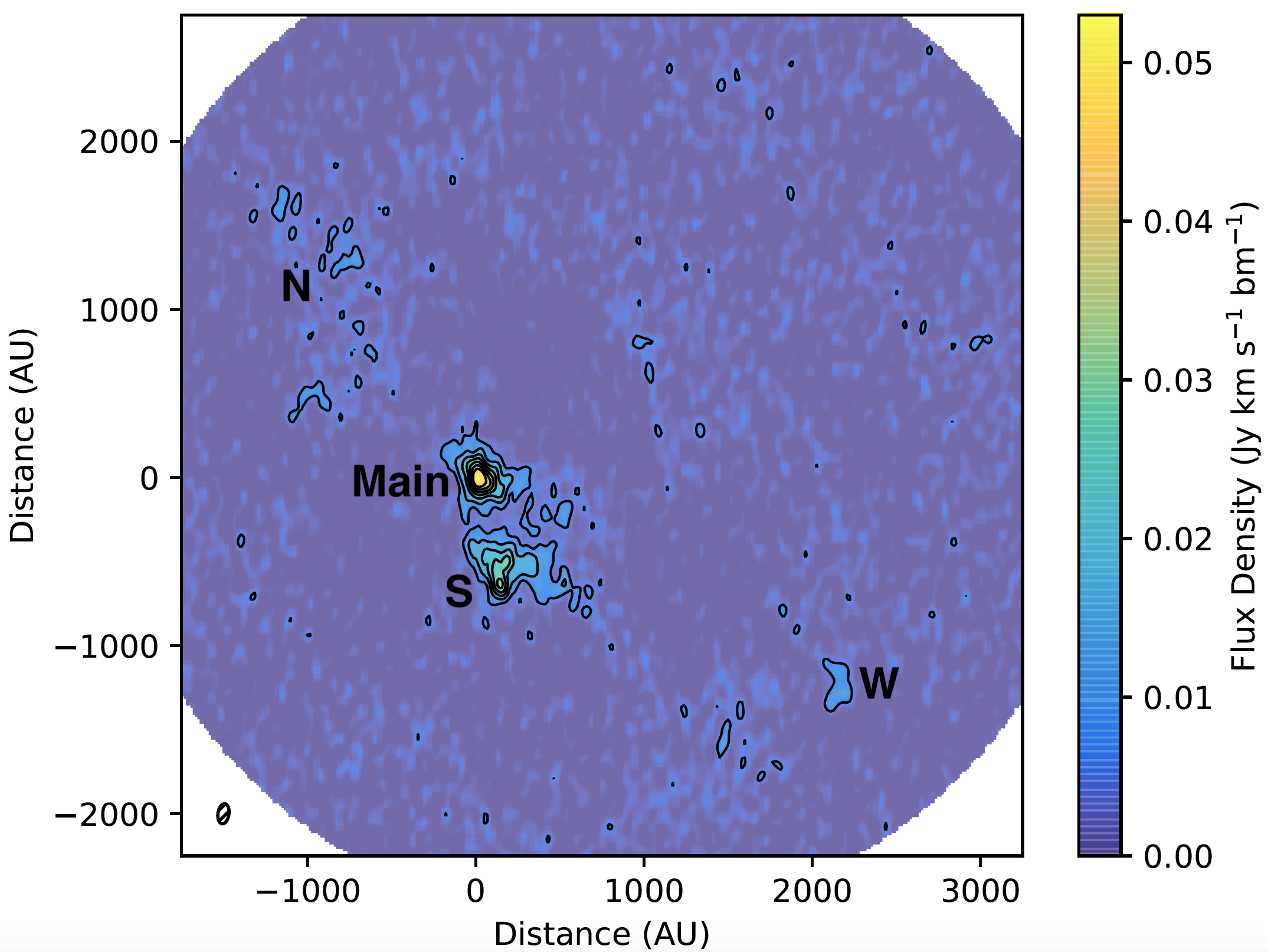}
	\caption{Primary beam corrected integrated intensity map (between 3.1 and 5.5~km~s$^{-1}$) for H$_2$CO transitions at 218.222~GHz. The continuum peak is at 0,0 au. Contours start at 3$\sigma$ (rms of 6 mJy/beam~km~s$^{-1}$) continue in intervals of 3$\sigma$ to a peak of 53 mJy/beam~km~s$^{-1}$. The beam size is 0.80$''$x0.43$''$.}
	\label{H2CO}
\end{figure}

H$_2$CO emission is detected toward the continuum as well as about 3$''$ ($\sim$600~au) to the south of the continuum source, about 16$''$ (3100~au) southwest of the continuum source, and 7.5$''$ (1500~au) northeast of the continuum source in all three detected transitions. We labeled these regions Main, south (S), west (W), and north (N) (See Figure~\ref{H2CO}). The emission from H$_2$CO in these three regions coincides with CS emission in all cases, but does not show the hourglass morphology toward the central source (Main) seen in CS. Cha-MMS1 W corresponds to the southwestern lobe of the CO outflow emission. There is emission detected from the H$_2$CO transition at 218.222~GHz about 10$a ''$ (1900~au) north-northeast of the continuum source, but the emission detected in this area from the other two transitions is very weak.  The higher energy H$_2$CO transition at 218.475 GHz (E$_{\mathrm{up}}$=68~K) shows very weak emission in each region (see Appendix A for spectra). 

Table~\ref{rotdTable} shows the rotational temperatures and column densities determined using the H$_2$CO transitions in a rotational diagram in the CASSIS software\footnote{http://cassis.irap.omp.eu \citep{Vastel2015}. CASSIS has been developed by IRAP-UPS/CNRS.} and the CDMS database. Rotational temperatures range from $\sim$25~K to 51~K with a temperature of 51.0~($\pm$36.2)~K toward the continuum source. The column densities are consistently $\sim 10^{13}$~cm$^{-2}$. The gas in these condensations appears to be sufficiently dense for the rotational temperature, gas temperature, and dust temperature to be equal (n$>10^{4.5}~cm^{-3}$ from \citet{Goldsmith2001}). In this case, based on the experimentally-measured sublimation temperature (144~K, \citet{Noble2012}), it is unlikely that the detected H$_2$CO originates in thermal desorption from warm dust. A gas-phase origin in reactions between oxygen atoms and CH$_3$ radicals is possible. 

\begin{table}[h]
	\centering
	\caption{H$_2$CO gas properties in different regions of Cha-MMS1 with errors in parentheses. Note that the fit for N was poor due to the low flux of the transition at 219475 MHz (E$_{\mathrm{up}}$ 68 K).}
	\label{rotdTable}
	\begin{tabular}{ccc}
		\hline\hline
		Region & T$_{rot}$             & N$_{col}$        \\
		& (K) & (cm$^{-2}$) \\
		\hline
		Main                 & 40.4 (+2.7, -2.4)   & 3.8 (0.4) $\times 10^{13}$ \\
		N                    & 73 (+101, -27)  & 2.8 (+0.08, -0.01) $\times 10^{13}$  \\
		S                    & 10.8 (+0.3, -0.2)  & 1.4 (0.1) $\times 10^{13}$ \\
		W                    & 36.5 (+7.0, -5.1) & 3.9 (+1.2, -0.9) $\times 10^{13}$ \\
		\hline
	\end{tabular}
\end{table}

\subsection{CH$_3$OH emission}
\label{ch3oh}
CH$_3$OH emission is detected towards the clumps associated with H$_2$CO emission to the northeast, south, and southwest of the continuum source, but not toward the continuum source itself. All CH$_3$OH emission is much weaker than in other detected species.  The column densities are consistent between regions at $10^{13}$-$10^{14}$ cm$^{-2}$ (Table~\ref{rotdTableCH3OH}). The upper limit for the column density of CH$_3$OH toward the main source was calculated assuming a T$_{rot}$ of 51~K (to match the temperature of H$_2$CO toward Main) and a line width of 0.7~km~s$^{-1}$. The inferred dust temperatures are significantly lower than the sublimation temperature of CH$_3$OH derived from experiments (128~K; \citet{Penteado2017}) and so the observed methanol is likely being produced by ice sputtering in shock waves.

\begin{table}[h]
	\centering
	\caption{CH$_3$OH gas properties in different regions of Cha-MMS1. * The rotational temperature of CH$_3$OH towards Main was assumed to be 51~K to mirror H$_2$CO toward Main and the column density is therefore an upper limit.}
	\label{rotdTableCH3OH}
	\begin{tabular}{ccc}
		\hline\hline
		Region & T$_{rot}$   & N$_{col}$        \\
		& (K) & (cm$^{-2}$) \\
		\hline
		Main & 51* & $<$1.2 $\times 10^{13}$  \\
		N  & 41.2 (30.8)     & 8.6 (6.4) $\times 10^{13}$  \\
		S  & 23.2 (2.3)    & 4.8 (0.8) $\times 10^{13}$  \\
		W & 15.7 (1.9)      & 1.1 (0.4) $\times 10^{14}$  \\
		\hline
	\end{tabular}
\end{table}

\begin{figure}[hb!]
	\centering
	\includegraphics[width=\hsize]{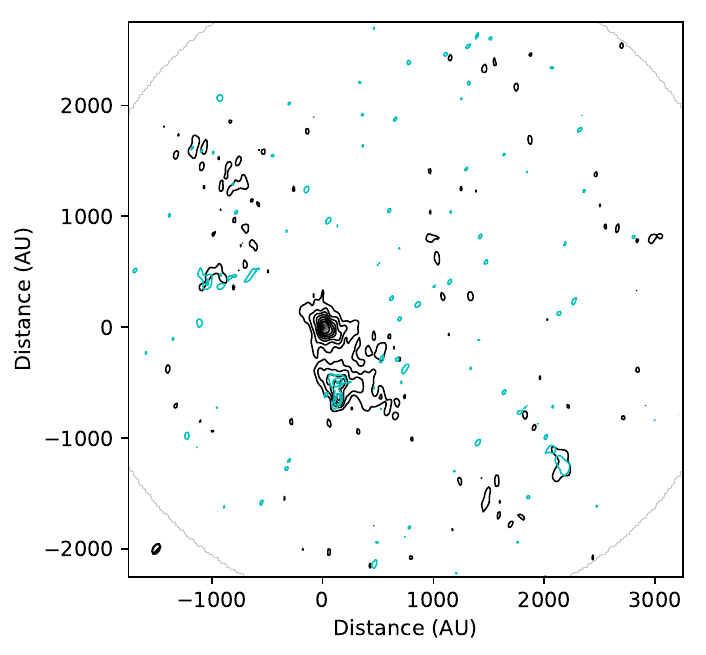}
	\caption{Primary beam corrected integrated intensity maps for H$_2$CO (218.222~GHz) and CH$_3$OH (241.791~GHz). Black contours show H$_2$CO as in Figure~\ref{H2CO}. Cyan CH$_3$OH contours (integrated between 3.6 and 5.4~km~s$^{-1}$) start at 3$\sigma$ (rms of 6.6 mJy/beam~km~s$^{-1}$) and continue in 1$\sigma$ intervals to a peak of 25 mJy/beam~km~s$^{-1}$. The continuum is shown in greyscale coincident with H$_2$CO emission, but not CH$_3$OH.}
	\label{methanolh2co}
\end{figure}



\section{Discussion} \label{sec:discuss}

\subsection{Physical origin of outflow emission}

The CO emission has multiple spatially overlapping components indicating that the outflow is nearly perpendicular to the line-of-sight. The shape of the red-shifted emission in CO and $^{13}$CO to the southwest of the continuum source implies that there is a dense clump of molecular gas diverting the outflowing material, which is traced in our observations by H$_2$CO and CH$_3$OH. This is also suggested in \citet{MariaJose2020} in maps of NH$_{2}$D. The alternative scenario is that this S-shaped emission and slower, more distant blue-shifted emission indicates a precessing outflow. The broad red-shifted emission to the northeast of the continuum may originate from a second source implying a tight binary system within the continuum. There is no corresponding blue-shifted broad component detected at similar velocities but it may have been disrupted by the molecular clump south of the continuum source.  The irregular velocity structure demonstrates that in the early stages of star formation, the outflow launching is chaotic. The clumpy structure observed in CS and H$_2$CO emission may be due to small ejections from the early stages of the stellar outflow or simply the clumpiness of the surrounding cloud. The underlying power source driving the outflow activity associated with Cha-MMS1 could be either a single precessing outflow or two outflows from a close binary system. Alternative scenarios can also produce outflow precession, for instance, the misalignment between the outflow axis (a proxy for angular momentum axis) and the magnetic field axis as discussed in \citet{Busch2020}, which does not require a binary. Similarly, a change in the angular momentum axis of the infalling material due to turbulent conditions can also produce shape distortions \citep{Matsumoto2017}.

\subsection{Status of Cha-MMS1}
While many FHSC candidates have been proposed over the past 20 years, most have now been identified as young class 0 objects, VeLLOs, or prestellar cores. Compared to the outflow properties of the remaining FHSC candidates with published outflow properties (G208.89-20.04 \citet{Dutta2022}, MC35-mm (I1535-NE) \citet{Fujishiro2020}, L1451-mm \citet{Pineda2011}, and GF 9-2 \citet{Furuya2019}) Cha-MMS1 has a similar maximum velocity range, generally lower dynamical time, and higher mechanical luminosity (by at least 2 orders of magnitude). The mechanical luminosity is also several orders of magnitude higher than the VeLLOs identified in \citet{Dunham2011}. The sources with the most similar outflow properties are the young stars in Barnard 1b \citep{Hirano2014}. Numerous complex molecular lines have been detected toward B1b-S \citep{Marcelino2018}, indicating that it is too evolved to be a FHSC. 
 
\subsubsection{Mechanical properties}

The mechanical outflow properties are several orders of magnitude lower than a 1 L$_{\sun}$ star. The mechanical properties we have derived for Cha-MMS1 are in line with the relationships between mass and luminosity and momentum rate and luminosity shown in \citet{Wu2004}.
The line of sight outflow properties we calculated using the CO (2-1) transition are generally similar to the values in \citet{Busch2020}, which were calculated using the CO (3-2) transition. They concluded that Cha-MMS1 was not a FHSC, though it was one of the youngest Class 0 objects ever detected. We find the kinetic time to be on the order of hundreds of years, rather than thousands, which results in our time-dependent values being an order of magnitude higher.  We agree with their assessment that the bent shape of the red-shifted emission in CO and $^{13}$CO to the southwest of the continuum source is likely due to the clump of molecular material there.


\subsubsection{Chemistry}
The presence of H$_2$CO without corresponding CH$_3$OH towards the main source is striking and attests to the young age of this source.  CH$_3$OH is also underabundant with respect to the CH$_3$OH/H$_2$CO ratio typically observed in the vicinity of low and high-mass protostars \citep{Floris2000}.
Using detected emission from three H$_2$CO transitions and three CH$_3$OH, we created rotational diagrams to determine the gas temperature and column density at the four positions described in the previous section: main, N, S and W. The resulting gas temperatures at Cha-MMS1 main and W are $\sim$50~K, indicating dust temperatures below those required to thermally desorb H$_2$CO (114~K) from icy dust grain mantles. The low dust temperature inferred from the temperature of CH$_3$OH is too cool for thermal desorption, which implies that this emission arises from ice sputtering from interactions with the out-flowing material, as CH$_3$OH is not easily made in the gas phase. The rotational temperatures for the emission at Cha-MMS1 S suggest dust temperatures lower than the sublimation temperature of both H$_2$CO and CH$_3$OH, therefore may also be the result of outflow shocks sputtering ices.




\section{Conclusions}
\label{sec:conc}
We observed Cha-MMS1 with ALMA in Band 6 at high angular resolution and detected $^{12}$CO and $^{13}$CO emission as well as CS, H$_2$CO, and CH$_3$OH. We find a weak outflow nearly perpendicular to the line of sight toward the northeast and southwest of the continuum source. The effects of the outflow are seen on the local molecular cloud through CS, H$_2$CO, and CH$_3$OH emission. A kink in the southwestern outflow corresponds with low rotational temperature ($\sim$20-30~K) emission from H$_2$CO and CH$_3$OH indicating a clump of molecular material being shocked by the outflow. 

\par Comparison of Cha-MMS1 with other low-luminosity protostars supports the view that it is not a viable FHSC. In fact, recent observations of SiO and CH$_3$OH in L1451-mm \citep{Wakelam2022} suggest that it too is not a FHSC. 
The weak outflow and relatively low temperatures associated with this object indicate that it is one of the youngest known Class 0 objects.




\acknowledgments
      \textit{{\large Acknowledgments:}}
      This paper makes use of the following ALMA data: ADS/JAO.ALMA 2013.1.01113.S. ALMA is a partnership of ESO (representing its member states), NSF (USA), and NINS (Japan), together with NRC (Canada) and NSC and ASIAA (Taiwan), in cooperation with the Republic of Chile. The Joint ALMA Observatory is operated by ESO, AUI/NRAO, and NAOJ. The National Radio Astronomy Observatory is a facility of the National Science Foundation operated under a cooperative agreement by Associated Universities, Inc. V. Allen's research for this article was supported by an appointment to the NASA Postdoctoral Program at the NASA Goddard Space Flight Center, administered by Universities Space Research Association under contract with NASA. V. Allen is currently supported by an NWO Veni grant number VI.Veni.202.135. V. Allen, M. Cordiner, and S.B. Charnley were supported by the NASA Planetary Science Division Internal Scientist Funding Program through the Fundamental Laboratory Research work package (FLaRe). 

\vspace{5mm}
\facilities{ALMA}
\clearpage

\appendix
\section{Spectra}
\begin{figure*}[!h]
	\centering
	\includegraphics[height=0.5\vsize]{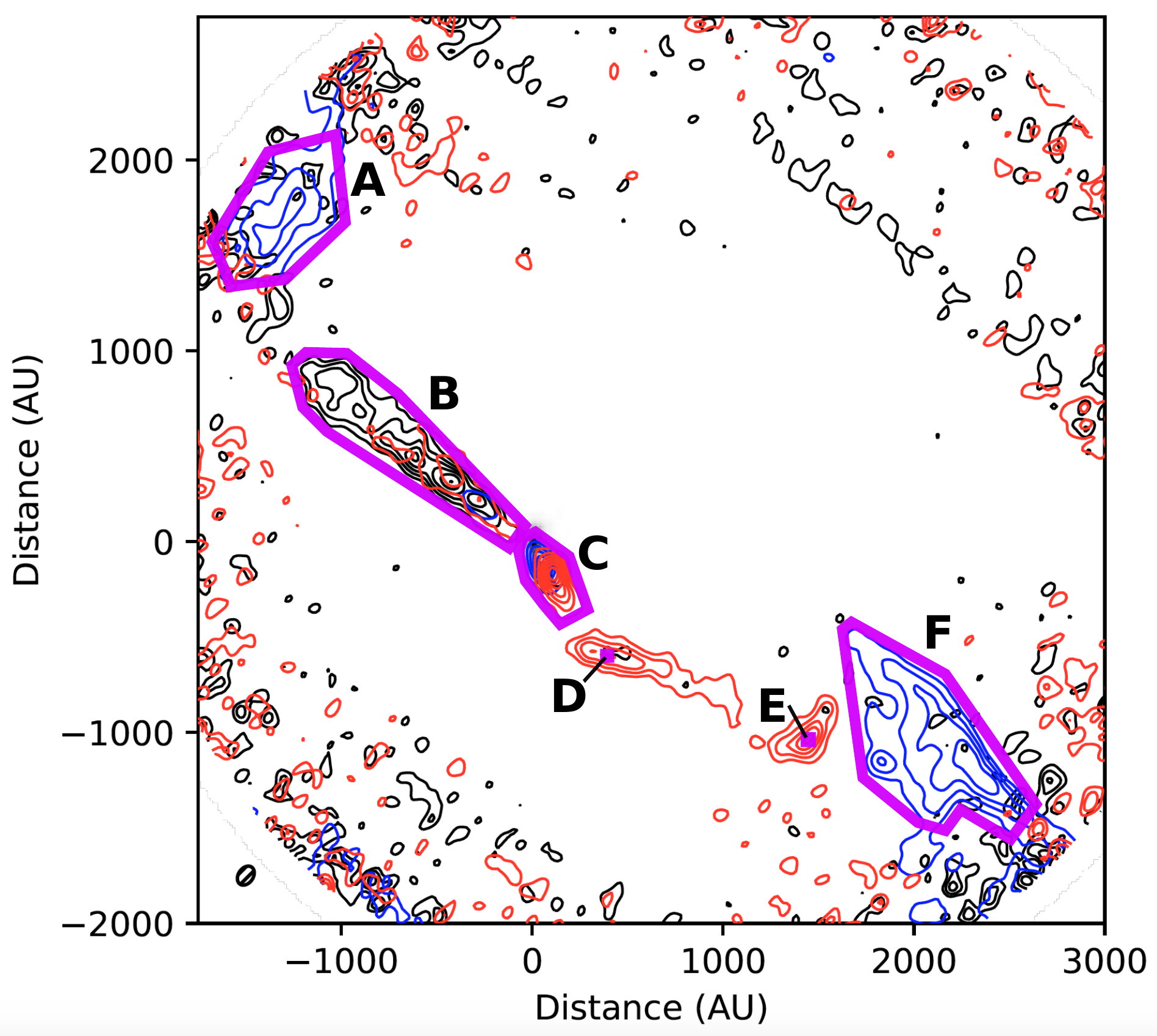}
	\caption{Integrated intensity maps of CO as in Figure~\ref{COallcontours} with sub-regions labeled A-F. All spectra are integrated over the area surrounded in magenta, except D and E which are taken from the emission peak.}
	\label{COlabels}
\end{figure*}
\begin{figure*}[ht!]
    \begin{minipage}{0.33\textwidth}
		\centering
		\includegraphics[width=\textwidth]{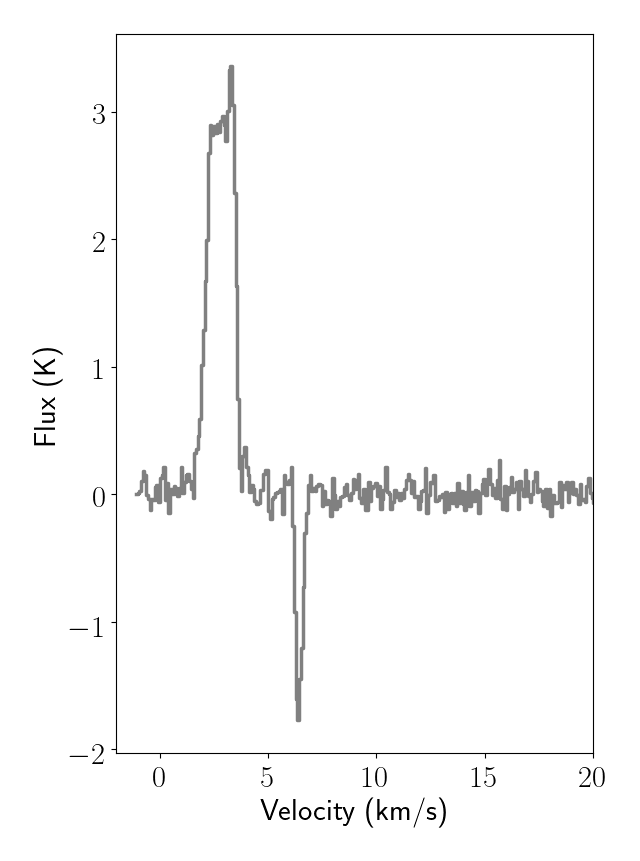}
    \end{minipage}  
    \begin{minipage}{0.33\textwidth}
		\centering
		\includegraphics[width=\textwidth]{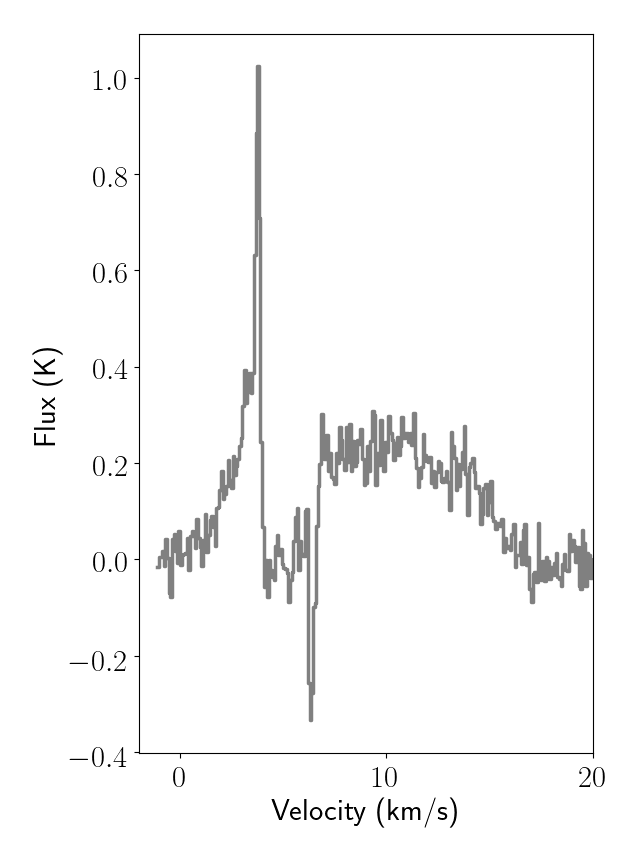}
    \end{minipage}
    \begin{minipage}{0.33\textwidth}
		\centering
		\includegraphics[width=\textwidth]{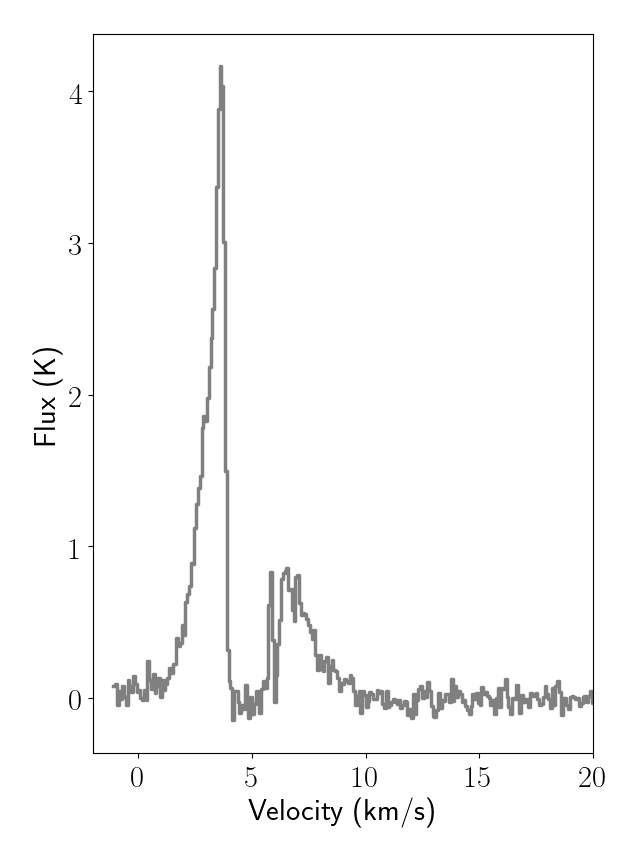}
    \end{minipage}
    \caption{CO spectra toward sub-regions A, B, and C (\textit{left to right}) from Figure \ref{COlabels}.}
	\label{cospec1}
\end{figure*}

\begin{figure*}[ht!]
    \begin{minipage}{0.30\textwidth}
		\centering
		\includegraphics[width=\textwidth]{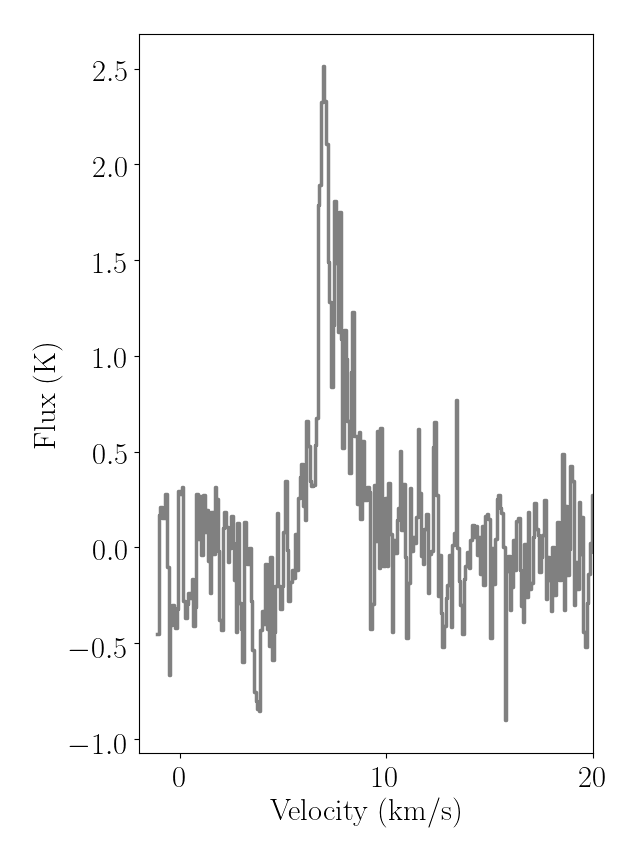}
    \end{minipage}
    \begin{minipage}{0.30\textwidth}
	\centering
		\includegraphics[width=\textwidth]{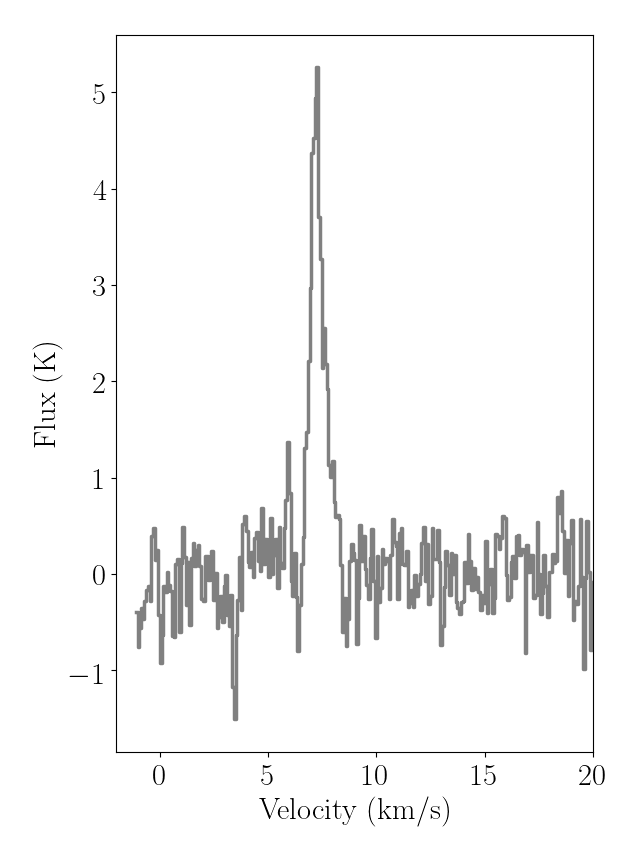}
    \end{minipage}
    \begin{minipage}{0.30\textwidth}
		\centering
		\includegraphics[width=\textwidth]{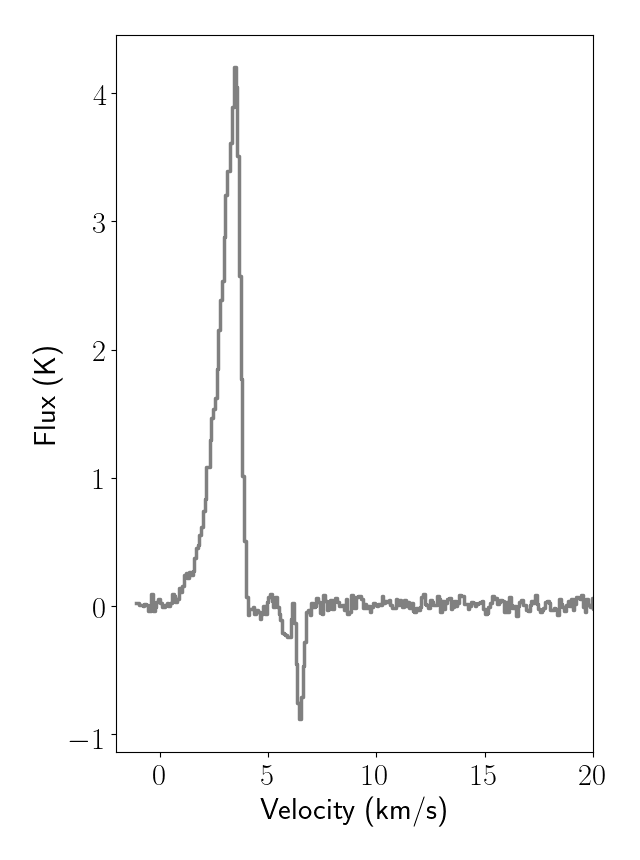}
    \end{minipage}
    \caption{CO spectra toward sub-regions D, E, and F (\textit{left to right}) from Figure \ref{COlabels}.}
	\label{cospec2}
\end{figure*}

\begin{figure*}[ht!]
    \begin{minipage}{0.23\textwidth}
		\centering
		\includegraphics[width=\textwidth]{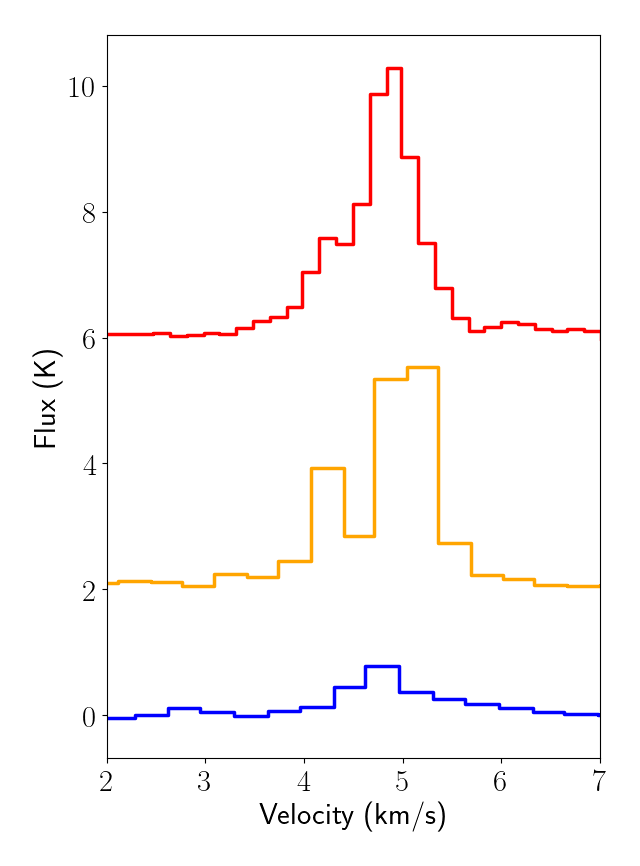}
    \end{minipage}
    \begin{minipage}{0.23\textwidth}
		\centering
		\includegraphics[width=\textwidth]{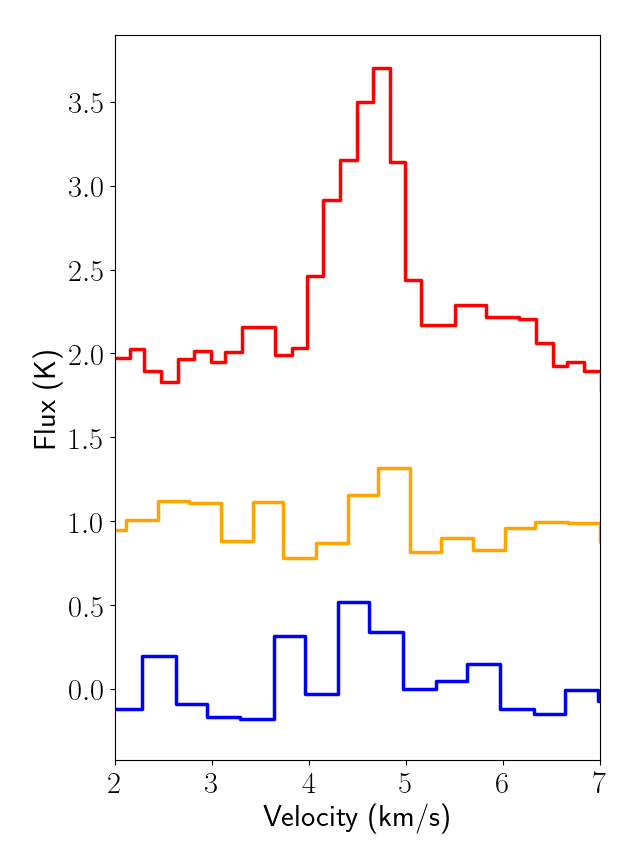}
    \end{minipage}
    \begin{minipage}{0.23\textwidth}
		\centering
		\includegraphics[width=\textwidth]{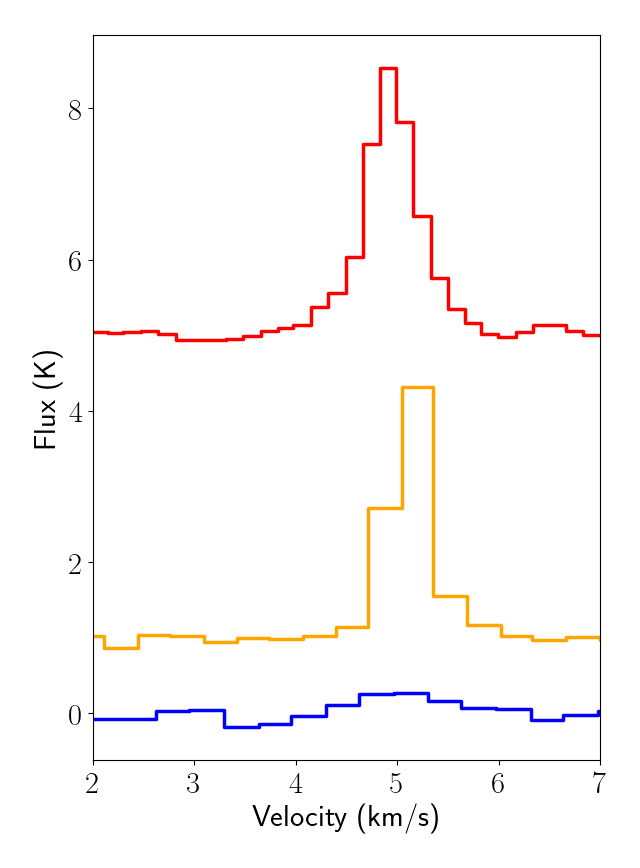}
    \end{minipage}
    \begin{minipage}{0.23\textwidth}
		\centering
		\includegraphics[width=\textwidth]{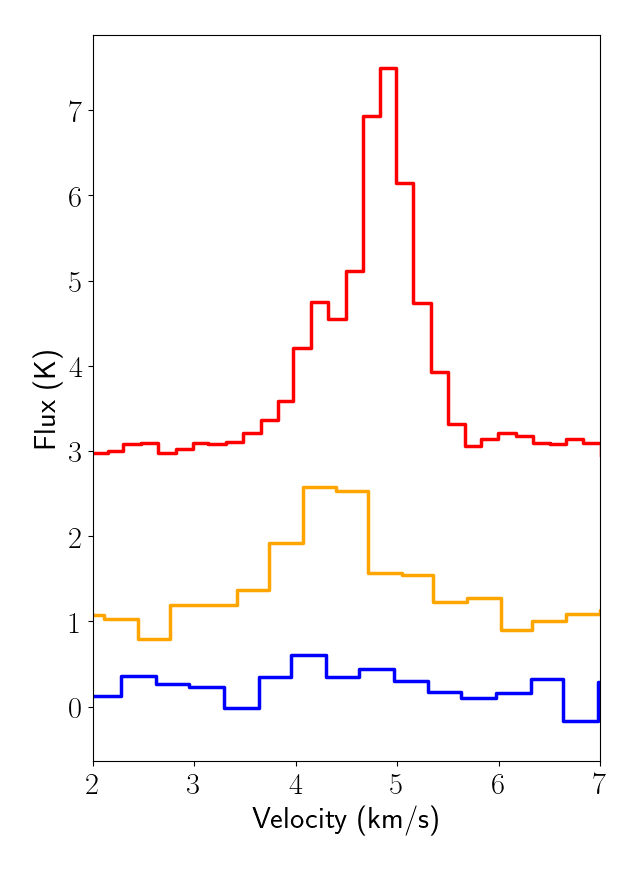}
    \end{minipage}
    \caption{H$_2$CO spectra toward each sub-region (\textit{left to right}): Main, N, S, and W. Red (top) corresponds to 3(0,3)-2(2,1) (218222 MHz), Yellow (middle) corresponds to 3(1,2)-2(1,1) (225698 MHz), Blue (bottom) corresponds to 3(0,3)-2(2,1) (218475 MHz).}
	\label{h2cospec}
\end{figure*}

\begin{figure*}[ht!]
    \begin{minipage}{0.23\textwidth}
	\centering
	\includegraphics[width=\textwidth]{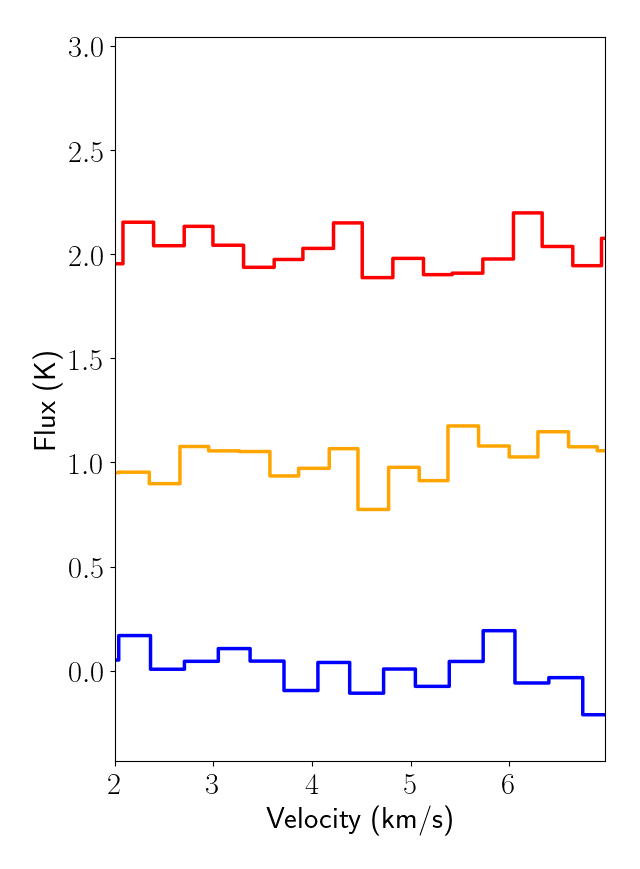}
    \end{minipage}
    \begin{minipage}{0.23\textwidth}
		\centering
		\includegraphics[width=\textwidth]{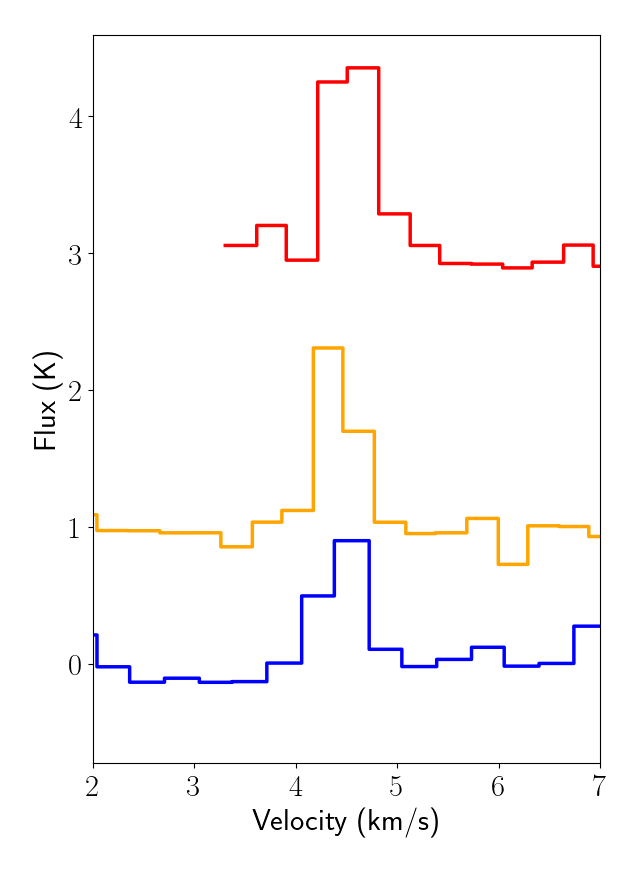}
    \end{minipage}
    \begin{minipage}{0.23\textwidth}
		\centering
		\includegraphics[width=\textwidth]{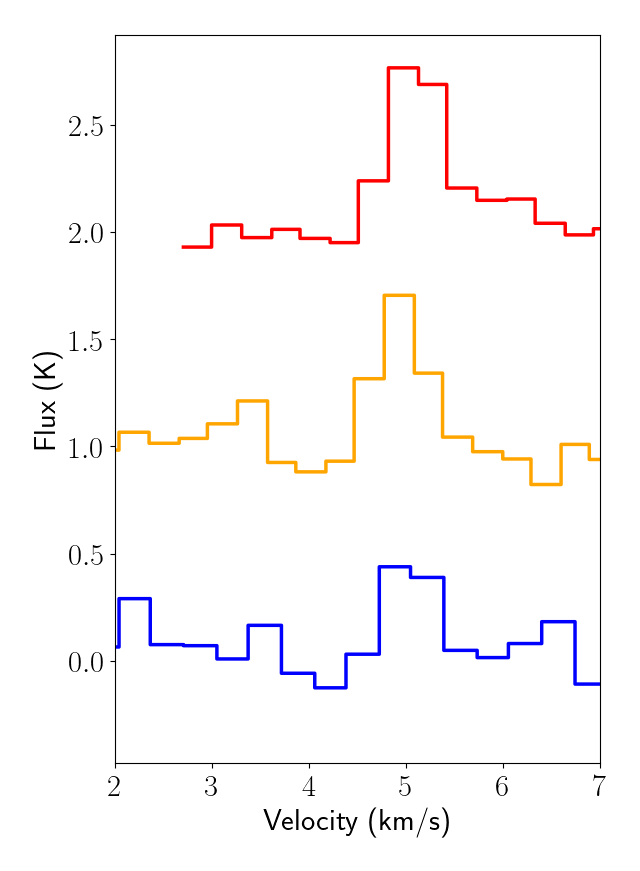}
    \end{minipage}
    \begin{minipage}{0.23\textwidth}
		\centering
		\includegraphics[width=\textwidth]{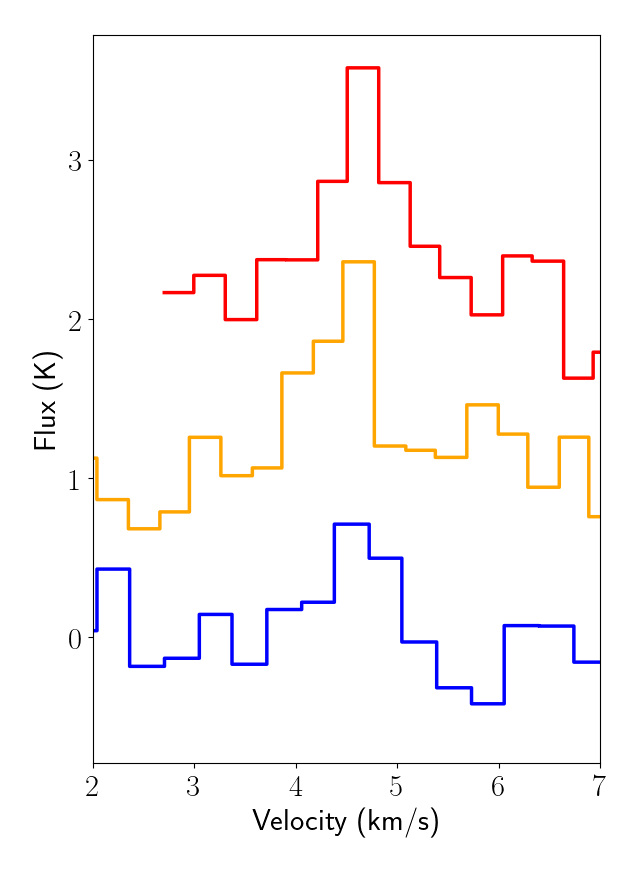}
    \end{minipage}
    \caption{CH$_3$OH spectra toward each sub-region (\textit{left to right}): Main, N, S, and W. Red (top) corresponds to 5(0,5)-4(0,4)++ (241791 MHz), Yellow (middle) corresponds to 5(-1,5)-4(-1,4) (241767 MHz), Blue (bottom) corresponds to 5(3,2)-4(3,1) (218440 MHz). CH$_3$OH emission is clearly not detected toward the main source.}
	\label{ch3ohspec}
\end{figure*}
\clearpage
\section{Additional integrated intensity maps}
\vspace{-0.5cm}

\begin{figure*}[ht!]
    \begin{minipage}{0.32\textwidth}
		\centering
		\includegraphics[width=\textwidth]{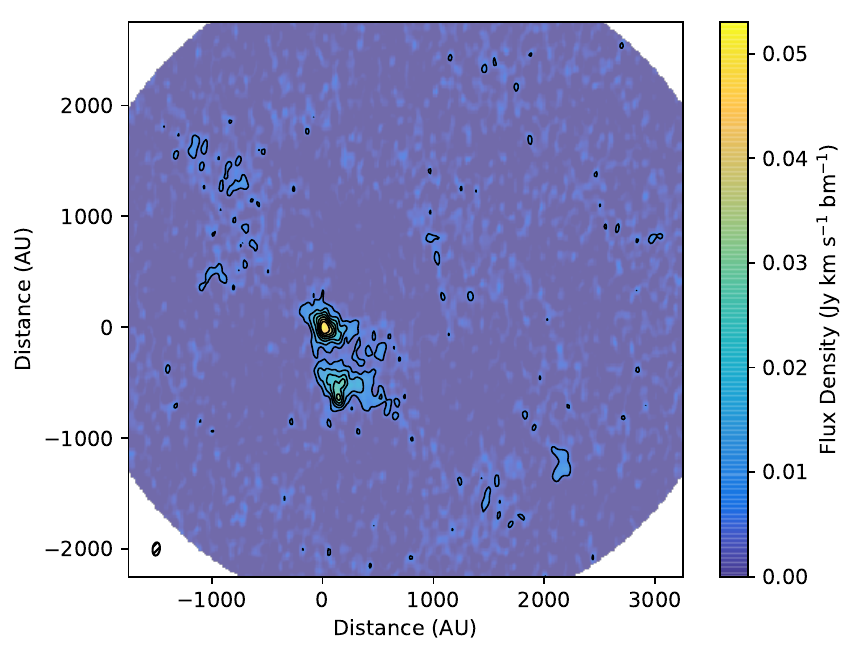}
    \end{minipage}
    \begin{minipage}{0.32\textwidth}
		\centering
		\includegraphics[width=\textwidth]{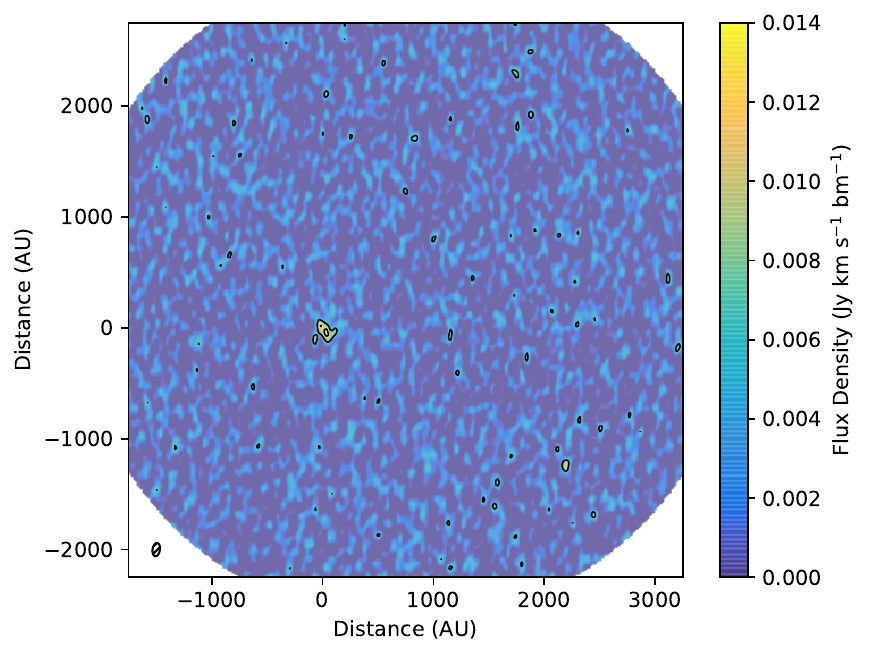}
    \end{minipage}
    \begin{minipage}{0.32\textwidth}
		\centering
		\includegraphics[width=\textwidth]{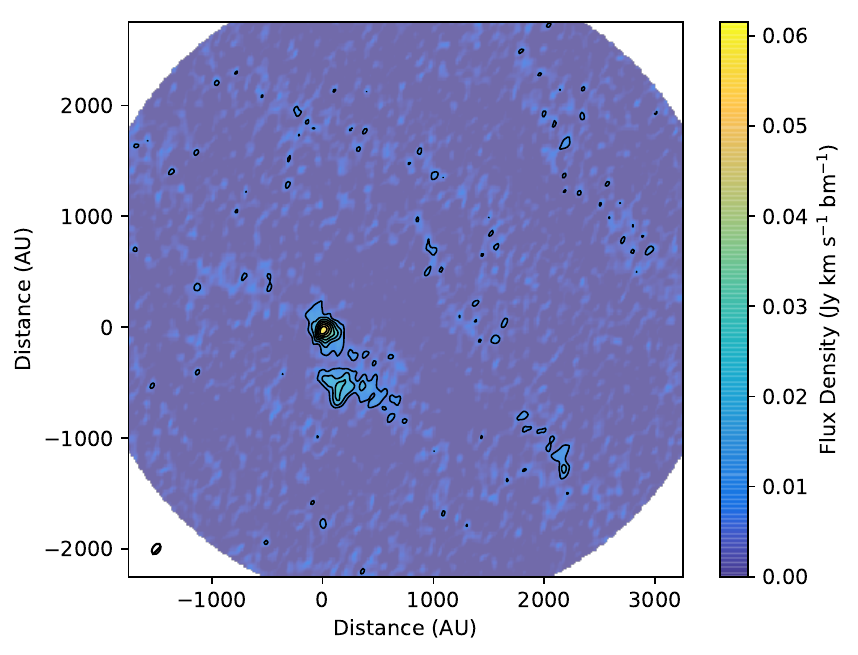}
    \end{minipage}
    \caption{Primary beam corrected integrated intensity maps for H$_2$CO transitions at \textit{(left)} 218.222~GHz (between 3.1 and 5.5~km~s$^{-1}$), \textit{(middle) }218.475~GHz (between 3.3 and 5.9~km~s$^{-1}$), and \textit{(right)} 225~GHz (between 3.7 and 6.0~km~s$^{-1}$).  The continuum peak is at 0,0 au.
    \textit{(left)} Contours start at 3$\sigma$ (6 mJy/beam~km~s$^{-1}$) continue in intervals of 3$\sigma$ to a peak of 53 mJy/beam~km~s$^{-1}$. The beam size is 0.80$''$x0.43$''$.
    \textit{(middle)} Contours start at 3$\sigma$ (6 mJy/beam~km~s$^{-1}$) continue in intervals of 3$\sigma$ to a peak of 14 mJy/beam~km~s$^{-1}$. The beam size is 0.81$''$x0.45$''$.
    \textit{(right)} Contours start at 3$\sigma$ (7.5 mJy/beam~km~s$^{-1}$) continue in intervals of 3$\sigma$ to a peak of 61 mJy/beam~km~s$^{-1}$. The beam size is 0.72$''$x0.44$''$.}
	\label{H2COcontours}
\end{figure*}

\begin{figure*}[ht!]
    \begin{minipage}{0.32\textwidth}
		\centering
		\includegraphics[width=\textwidth]{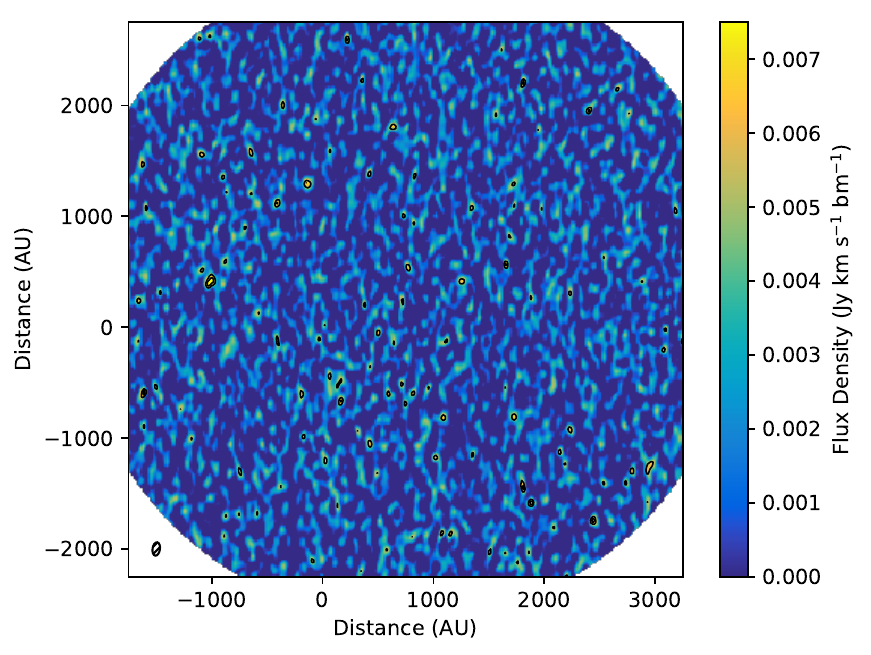}
    \end{minipage}
    \begin{minipage}{0.32\textwidth}
		\centering
		\includegraphics[width=\textwidth]{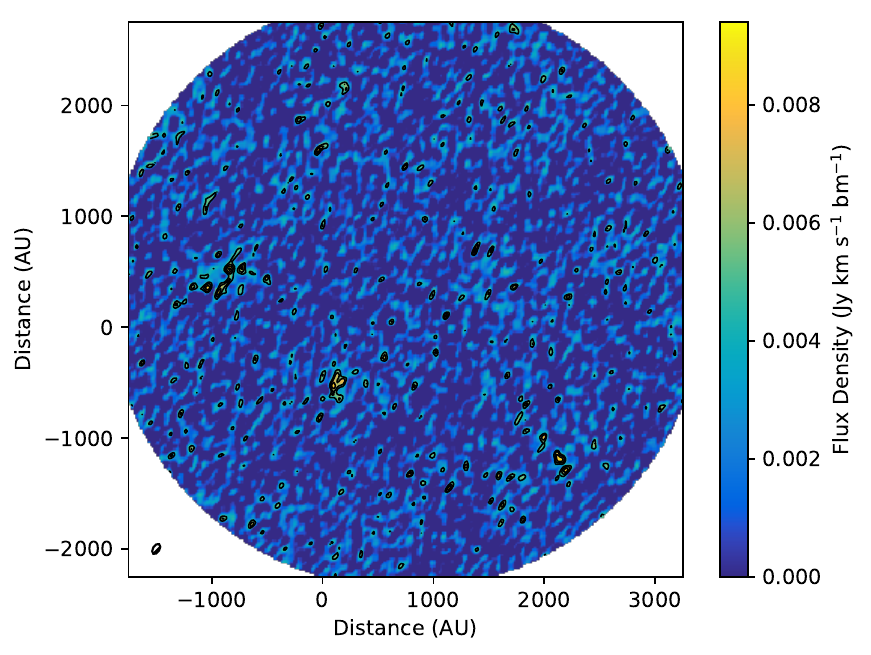}
    \end{minipage}
    \begin{minipage}{0.32\textwidth}
		\centering
		\includegraphics[width=\textwidth]{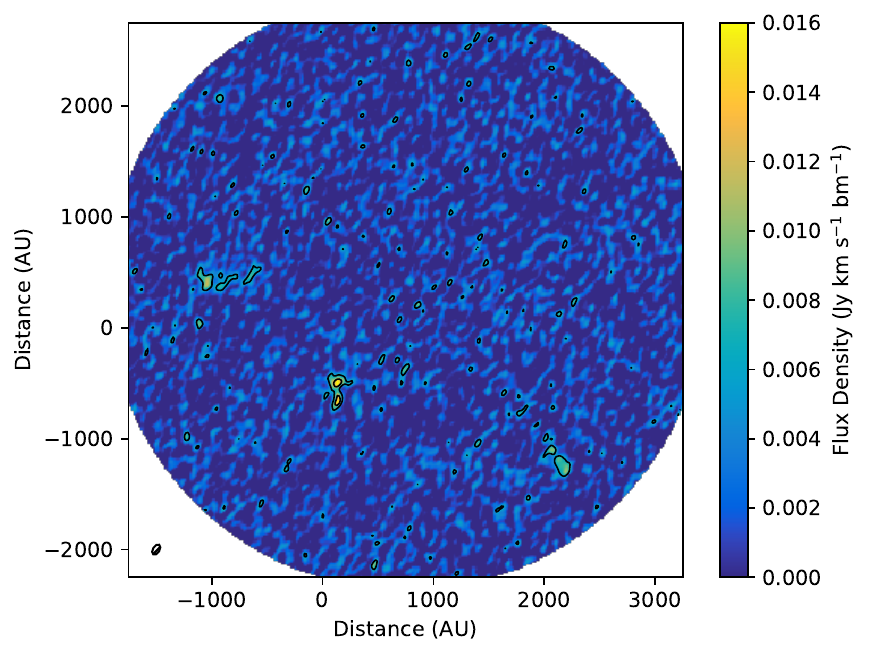}
    \end{minipage}
    \caption{Primary beam corrected integrated intensity maps for CH$_3$OH transitions at\textit{ (left)} 218.440~GHz (between 4.0 and 5.7~km~s$^{-1}$), \textit{(middle)} 241.767~GHz (between 3.9 and 6.0~km~s$^{-1}$), and \textit{(right)} 241.791~GHz (between 3.6 and 5.4~km~s$^{-1}$). The continuum peak is at 0,0 au.
    \textit{(left)} Contours start at 3$\sigma$ (5.4 mJy/beam) continue in intervals of 1$\sigma$ to a peak of 7 mJy/beam. The beam size is 0.81$''$x0.45$''$.
    \textit{(middle)} Contours start at 3$\sigma$ (4.5 mJy/beam) continue in intervals of 1$\sigma$ to a peak of 9 mJy/beam. The beam size is 0.68$''$x0.40$''$.
    \textit{(right)} Contours start at 3$\sigma$ (6.0 mJy/beam) continue in intervals of 3$\sigma$ to a peak of 16 mJy/beam. The beam size is 0.68$''$x0.40$''$.}
	\label{CH3OHcontours}
\end{figure*}

\end{document}